\documentclass[%
 reprint,
 amsmath,amssymb,
 aps,
prd,
]{revtex4-2}

\usepackage{aas_macros}
\usepackage{graphicx}
\usepackage{dcolumn}
\usepackage{bm}
\usepackage{xcolor}
\usepackage{ulem}[normalem]
\usepackage{amssymb}
\usepackage{hyperref}

\begin{document}



\title{A low viscosity relatively thick twisted disk in a supermassive binary
black hole as a potential model of OJ 287}
\author{{{Viacheslav}} V. Zhuravlev}
\email{v.jouravlev@gmail.com}
\affiliation{Sternberg Astronomical Institute, Lomonosov Moscow State University, Universitetskij pr., 13, Moscow 119234, Russia }%
\author{Pavel B. Ivanov}
\email{pavel000astrophysics@gmail.com}
\affiliation{%
P.N. Lebedev Physical Institute, 53 Leninsky Prospect, Moscow, 119991, Russia
}%

\begin{abstract}
In this Paper we consider twisted accretion disks in supermassive binary black hole by analytical and numerical means.  It is assumed that the disk orbiting around the more massive rotating component and that the disk rings are 
inclined with respect to the orbital plane.
We use orbital parameters of the binary, which are often employed in
the so-called precessing massive (PM) model of the activity of the well-known blazar OJ 287. In particular,
the orbit is assumed to have a large eccentricity $\sim 0.7$, and the ratio of mass of the secondary component to the
primary is $\sim 10^{-2}$.

Unlike our previous investigation of a similar problem, here we consider disks with both small and relatively large
relative thicknesses $\delta=h/r$, where $h$ is the disk's height at a typical radius $r$,
as well as a range of values of the viscosity parameter, $\alpha$, including the cases when $\alpha \lesssim
\delta$.
The influence of the secondary black hole
on the disk is treated using the formalism proposed by us elsewhere, which
is based on the double averaging procedure of the secondary's gravitational field.

Similar to our previous results, we find that the twisted disk relatively quickly relaxes to a quasi-stationary
state in the frame precessing with the Lense-Thirring frequency of the orbit. However, its shape is qualitatively
different from that corresponding to the case of $\delta=10^{-3}$ and $\alpha=0.1$ considered in our previous work.

When $\delta \sim 10^{-3}$ the shape of the disk of any viscosity is largely determined by
the resonance between a forcing frequency associated with the presence of the secondary and the Lense-Thirring
frequency of a particular disk ring calculated in the precessing frame. In addition to the effect determining 
the shape of the $\alpha=0.1$ disk
considered by us earlier, we find the new effect of generation of a twisting spiral wave near the resonance in a disk with $\alpha \le 2\cdot 10^{-2}$.
We propose an analytic theory of it, which is in quite good agreement with our
numerical results. For all cases, typical disk's inclinations with respect to the equatorial
plane are of the order of or larger than the orbital inclination. This leads to multiple crossings of the
orbit with the disk per one orbital period, which contradicts the PM model, where only two crossings per
orbital period are required.

When $\delta \gtrsim 0.1$, a typical disk's inclination within the orbit of the binary turns out to be smaller
than that of the orbit. We provide qualitative analytical analysis, which confirms this conclusion. In this case
 there are only two crossings of the orbit with the disk per one orbital period. On the other hand, qualitative estimates
 allow us to suggest that the additional heating of the disk gas by the secondary-disk collisions may result in
 $\delta \sim 0.1$. Thus, we suggest that in the framework of PM model of OJ 287 the disk should be relatively thick,
 with $\delta \sim 0.1$. This could lead to a modification of the theoretical spectrum of the source.

\end{abstract}

\maketitle

\onecolumngrid
\section{Introduction}
The famous source of quasi-periodic optical emission OJ 287 is widely considered as the best supermassive
black hole binary candidate. Perhaps the most popular model of its activity considers two supermassive
black holes of unequal masses and a planar accretion disk around the most massive black hole, called
hereafter the primary.  Since the orbital plane of the black holes and the disk midplane are assumed to be misaligned, 
the less massive black hole
called hereafter the secondary hits the disk twice per orbital period. These collisions produce some
shocked hot gas, whose emission is assumed to be responsible for the observed luminosity outbursts of OJ 287,
see e.g. \cite{Sil}, \cite{LV}, \cite{Valt}. An important ingredient of this model is an assumption that
certain details of the luminosity curve can be explained by the Einstein apsidal precession of the
orbit (see \cite{LV}), which allows one to fix the main parameters of the system. In particular, it turns
out that the model requires a very massive primary of mass $M\approx 2\cdot 10^{10}M_{\odot}$ and much
less massive secondary of mass $m\approx 1.5\cdot 10^{8}M_{\odot}$, while the orbital eccentricity, $e$,
needs to be rather large, $e\approx 0.7$. For these reasons, this model of activity of OJ 287 is sometimes
called the 'massive precessing' (PM) model.

Recently, some authors questioned the validity of the PM model. In particular, in \cite{Kom1} the authors
pointed out that the outburst predicted by the PM model to happen in December 2021 was, in fact, absent.
From that, the conclusion that the necessity of a highly precessing orbit, which leads to the exceptionally
large mass of the primary was made. However, the authors of \cite{Valt2} later modified the PM model
in such a way that the outburst time was shifted to July-August 2020, when the possibility to observe
it from Earth was absent.

Another point of criticism was raised in our paper \cite{IZ}. In this paper it was pointed out that
if the accretion disk is assumed to be the standard razor-thin $\alpha$-disk with its height to radius
ratio $\delta=10^{-3}$ and the $\alpha$ parameter is relatively large, $\alpha=0.1$, the disk takes a
non-trivial quasi-stationary shape in the reference frame precessing with the Lense-Thirring frequency
of the binary. This shape is illustrated below for twice as large $\delta$ and $\alpha$, 
see the bottom-left panel of Fig. \ref{Fig8} for a qualitative comparison with the results of \cite{IZ} The disk
shape is twisted near some particular radius, $r_r$, and a typical disk's inclination angle with respect to the
primary equatorial plane, $\beta(r=r_r)$, is of the order of the binary inclination $\beta_b$. It was shown
that the secondary crosses such a disk many times per orbital period (typically, $5-6$ times), which is clearly
incompatible with the requirement of the PM model that the number of crossings per one orbital period is two as
in the flat disk case.

The physical reason for such a strong response is determined by the presence of a resonance in the system.
Namely, in the formalism employed in \cite{IZ}, the action of the orbit on the disk was described in terms
of two numerically calculated frequencies $\Omega_1(r)$ and $\Omega_2(r)$, see eq. (\ref{e17c}) below.
The resonance takes place when the forcing frequency $\Omega_1$ is equal to the Lense-Thirring frequency
of the disk ring at this radius calculated in the precessing frame, see eq. (\ref{equ2}) below. At radii
$r\sim r_r$, the secular influence of the secondary gravitational field
on the disk is especially strong. It turns out that $r_r$
lies between the orbit periastron $r_p$ and apoastron $r_a$
in the case of the orbital parameters used in the PM model, $r_p < r_r < r_a$.
Therefore, the disk turns out to be strongly twisted on the path of the secondary, which explains
the effect of multiple disk's crossings in this model.

In this Paper, we extend the results of \cite{IZ} by considering disks with different $\delta$ and $\alpha$
using both analytical and numerical approaches. In our numerical work we use the model of interaction
of the binary and the disk based on the double averaging procedure, which was put forward in \cite{IZ}.
In this model the disk tilt and twist are described in linear approximation.  We also neglect self-gravity of the disk, see e.g.  \cite{LP} for a discussion of self-gravitating twisted disks.
The binary orbital parameters remain to be the same as
in \cite{IZ} as well as the mass ratio $q=m/M$. These values correspond to those used in the standard variant
of PM model, see e.g., \cite{Valt}. We also fix the value of the primary black hole rotational parameter, $\chi$,
to be $0.5$ and the orbital inclination angle $\beta_b$ is fixed to be $0.1$. We assume that our results remain
qualitatively the same for other values of  $\chi \sim 1$. We briefly analyze the applicability of our results to the
case of larger $\beta_b$ at the end of this Paper, 
and find that they are likely to remain qualitatively the same when $\delta$ is large enough. 

We find that that in all considered cases the disk relaxes to a quasi-stationary state in the precessing frame. However,
its shape is drastically different for 'thin' disks with $\delta \sim 10^{-3}$ and 'relatively thick disks' with $\delta
\gtrsim 0.1$.

When the disk is thin, and viscosity parameter $\alpha \ll 1$, the influence of the resonance on the disk structure is
even more pronounced than in the case of relatively large alpha considered in \cite{IZ}. In addition to the effect
of a sharp rise of the disk inclination angle $\beta$ when $r \rightarrow r_r$, the disk region in the vicinity
of resonance generates waves of disk twist with a characteristic wavelength much smaller than $r_r$.
These waves are stationary in the precessing frame. Although disks with such parameters cannot be viable models
of OJ 287, we consider the mechanism of the generation of these waves in some detail, taking into account that
it could operate in other systems. From the mathematical point of view, this mechanism is analogous to the
well known mechanism of generation of density waves at Lindblad resonances (e.g., \cite{GT}) and we develop
a similar theory to compare it with numerical results below. We note, however, that physically our effect
is different from the effect of density wave launching. While in our case it is determined by the resonance 
between secular quantities doubly averaged over the mean anomalies of the binary and a particular gas element of 
the disk, the Lindblad resonance is an effective mean-motion resonance between
the perturber and the disk gas.

When the disk is sufficiently thick, its shape differs quite significantly from the thin disk case.
Our numerical results suggest that when $\delta \gtrsim 0.05$, the disk inclination is smaller than $\beta_b$ within
the orbit (i.e., when $r_p < r < r_a$), see Figs. \ref{Fig6}
and \ref{Fig7} below. Also, the distribution of $\beta(r)$ within the orbit is quite featureless in stark contrast with
the previous case. In the latter case, $\beta$ slowly decreases with $r$, with a characteristic scale of this decrease growing
with $\delta$. Also, typical values of $\beta$ within the orbit decrease when $\delta$ grows.
We argue that in the case of such a thick disk, the resonance is not important, and we develop a qualitative analytic
approach to estimate $\beta (r)$, which agrees with our numerical results by a factor of $2-3$. Thus, the shape of
the thicker disks are much more appropriate for the PM model, and we demonstrate below that there are indeed only two disk crossings
by the secondary per one orbital period in this case.

The standard $\alpha$-theory of black hole accretion disks
(\cite{Sh} and \cite{SS}) predicts that accretion disks around supermassive black holes should
be razor-thin, with $\delta \sim 10^{3}$, see e.g. \cite{IIN}. We point out at the end of the Paper that the collisions of
the secondary with the disk may provide an additional source of the disk's heating and, therefore, enlarge $\delta$ to some extent.
Our preliminary estimates made in Section \ref{estimates} suggest that the disk at the scales order of interest
may be in the radiation dominated state.
Assuming that the thermal instability operates in the disk, it is then reasonable to assume
that $\delta \sim 1$, see e.g., \cite{Marek}. Under the opposite assumption that the thermal instability
is absent (see e.g., \cite{Blaes} and references therein), our estimates show that $\delta$ could be of the order of
$0.1$. Therefore, the requirement that the disk should be relatively thick to have only two black hole-disk collisions
per one orbital period may be consistent with the thick disk models, since the disks may quite significantly increase
their height due to additional gas heating determined by these collisions.

The structure of the Paper is as follows. In the following Section \ref{basic} we introduce our main definitions
and discuss the problem setting in a more quantitative way. In Section \ref{analyt} we develop an analytic WKBJ
(Wentzel-Kramers-Brillouin-Jeffreys) theory of propagation of the waves of twist and tilt as well as their
generation near the resonance in the thin disk case. In the same Section, we consider the thick
disk case and provide some simple
qualitative approach, which allows one to estimate a typical value of $\beta$ within the orbit up to a factor order
of unity. In Section \ref{numerical} we discuss the comparison between the analytic approaches and the numerical
results. We evaluate typical numbers of the disk crossings for the thin and thick disk cases and provide some three-dimensional illustrations
of geometrical shape of the thin and thick disks considered in this Paper comparing them to the shape of the disk considered in \cite{IZ}.
Finally, in Section \ref{discussion} we discuss the applicability of our results to the case of larger $\beta_b$ (see
Section \ref{nonlinear}),
the potential role of the secondary-disk collisions in heating of the disk and roughly estimate a possible value
of $\delta$, see Section \ref{estimates}.  We also discuss potential extensions of our work and make final conclusions.


\bigskip
\section{Basic definitions and notations}
\label{basic}
\subsection{Main quantities associated with the accretion disk}

As was discussed by many authors (see e.g. \cite{PP, II, DI, LP, ZI}, and references therein) when the disk's tilt is considered to be formally
small, a twisted disk model can be split into two parts: a background flat disk model and a model describing the evolution of the disk's tilt and twist. 
As a background model, we consider an $\alpha$-disk model having a constant ratio of disk's half-thickness
$h$ to the radial distance $r$, $\delta=h/r$, far from the black hole. Unlike the standard
\cite{Sh,SS} $\alpha$-model we treat $\delta$ as a free parameter.
This allows us
to describe a thicker disk than a razor-thin disk with $\delta \sim 10^{-3}$ which is predicted
by the standard model for accretion disks around supermassive black holes, see e.g., \cite{IIN}. Thus, our models of the background disk are determined by
two parameters - $\delta$ and the viscosity parameter $\alpha$.

The important characteristic frequencies characterizing the background disk model are
the Keplerian angular frequency $\Omega_{K}(r)$, a 'sound' frequency
associated with the possibility of propagation of tilt in
a low viscosity accretion disk with the speed two times smaller than its sound
speed (see \cite{LP}),
and two frequencies determined by relativistic effects of the Einstein apsidal precession,
$\Omega_{E}$, and the Lense-Thirring nodal
precession of nearly circular orbits of the disk gas, $\Omega_{LT}$.
Explicitly, we have
\begin{align}
\Omega_K=\sqrt{GM\over r^3}, \quad
\Omega_s={\delta \over 2}\Omega_K, \quad \Omega_E={3r_g\over r}\Omega_{K}, \quad
\Omega_{LT}=2\chi {({r_g\over r})^{3/2}}\Omega_K,
\label{d1}
\end{align}
where $G$ is the gravitational constant, $c$ is the speed of light,
\begin{align}
r_g={GM\over c^2},
\label{d2}
\end{align}
and $\chi$ is dimensionless rotational parameter of the central black hole, $|\chi| < 1$.

\subsubsection{The twisted disk model}

Following \cite{IZ}, for a numerical work, we use a time dependent, fully relativistic semi-analytic twisted disk model proposed in \cite{ZI}. In this model, the black hole rotational parameter
is formally small, and the flat disk model formally coincides with the Novikov-Thorne
extension of the Shakura-Sunyaev $\alpha-$model for a Schwarzschild black hole of mass $M$,
see \cite{NT}.
Accordingly, the radius of marginally stable circular orbit, $r_{ms}$, is always
equal to $6r_{g}$, and we consider radii $r > 6r_{g}$ in all our numerical simulations.
This model was tested against fully
relativistic magneto-hydrodynamical simulations in \cite{Danila, Zhur} and used to describe twisted disk 
dynamics after tidal disruption of a star by a black hole, see  \cite{IZP}.
We would like to
stress that, since characteristic scales associated with our model are larger than $r_{ms}$,
it is expected that our numerical model is qualitatively correct even in the case when
$|\chi| \sim 1$.

The main dynamical variables of this model are the inclination angle $\beta(t,r)$ and nodal angle $\gamma(t,r)$ of a disk's ring situated at a particular radius $r$ in a given moment
of time $t$ defined with respect to the primary black hole equatorial plane.
We also use the combination $W(r,t)=\beta e^{i\gamma}$, which is especially convenient for a description of twisted disk's
with a formally small disk's inclination $\beta$. The dynamical model is linear in $\beta$.

The variants of the numerical model considered in \cite{ZI, IZ} and in \cite{Danila, Zhur}
are slightly different due to different assumptions about the vertical structure of the flat
model. While in the former case it is assumed that the disk's density has a Gaussian distribution with height, in the latter case a polytropic falloff was considered. In this study we use the Gaussian law, corresponding to a locally isothermal disk,
$\rho (r,z)=\rho_{0}(r)\exp (-{z^2\over 2h^2})$, where the disk half-thickness $h=D(r)\delta r$, $\delta$ is assumed to be constant, and $D(r)$ is the usual relativistic factor, which formally ensures that this disk surface density tends to zero when $r\rightarrow r_{ms}$,
see \cite{NT} and \cite{PT}. We note that this requirement isn't physically
justified for thicker disks considered in this study, but, nevertheless, we use it for the disks with any considered
$\delta$, since it contributes to numerical stability of our model at $r \sim r_{ms}$.

\subsection{Orbital parameters of the black binary black hole and the main relations
determined by the presence of the secondary}
In this Paper we use the same model of the SBBH as in \cite{IZ}.
Namely, following \cite{Valt}
we fix the primary and secondary black hole masses $M$ and $m$
equal to $2\cdot 10^{10}M_{\odot}$ and  $1.5\cdot 10^{8}M_{\odot}$, respectively,
with the corresponding mass ratio $q=m/M=7.5\cdot 10^{-3}$. The binary semi-major axis, $a$,
is fixed to be equal to $60r_g$.
It gives the orbital period in the observer's frame,  $P_{obs}\approx 12yr$ as suggested by the application of the PM
model of OJ 287
to observations. The binary eccentricity, $e$, is equal to $0.7$ throughout the  text.
Accordingly, the orbital periastron $r_p=(1-e)a$ and the orbital apoastron $r_a=(1+e)a$
are equal to $18r_g$ and $102r_g$, respectively.

The binary's orbit is assumed to be inclined with respect to the equatorial plane of the central black
hole at an inclination angle $\beta_b$. Other angles characterizing the orientation
of the orbit are the nodal angle $\gamma_b$, where the line of nodes is defined with respect
to the equatorial plane, and the apsidal angle $\Psi_b$ defined as the angle in the
orbital plane between the line of nodes and the direction to the orbital periastron, see
\cite{II, ZI, IZ} for a detailed description of the coordinate systems used in this study.


The important characteristic frequencies associated with the SBBH's orbit are
the Keplerian frequency $\Omega_{K}^b$, the frequency of Einstein
apsidal precession $\Omega^{b}_{E}$ and the frequency of
Lense-Thirring precession of the orbital plane:
\begin{equation}
\Omega_{K}^b=\sqrt{GM\over a^3}, \quad \Omega^{b}_{E}={3r_g\over \epsilon^2 a}\Omega^b_K,
\quad \Omega^{b}_{LT}={2\chi G^2M^2\over c^3\epsilon^3 a^3},
\label{frecb}
\end{equation}
$\epsilon=\sqrt{1-e^2}$ and $\chi$ is the rotational parameter
of the central black hole.
We have $\Omega^{b}_E/\Omega^b_K \approx  10^{-1}$ and
$\Omega^b_{LT}/\Omega^b_K\approx 10^{-2}\chi$ for the chosen parameters of the binary. Clearly, we have
$\Psi_b=\Omega_E t +\Psi_b(0)$ and $\gamma_b=\Omega_{LT}^{b}t +\gamma_b (0)$, where
$\Psi_b(0)$ and $\gamma_b(0)$ are initial values of the angles. In what follows, we set
the initial values to zero.

The gravitational influence of the binary on gas elements of the disk is treated
in the framework of the model developed in \cite{IZ}. In this model, the double averaging over
the mean anomalies of the orbits of the binary and a particular gas element in the disk
is made, the inclination $\beta_b$ is assumed to be small and the gravitational influence
of the secondary on gas elements  situated in the vicinity of the points, where
the secondary crosses the disk is neglected.
Namely, it is assumed that the disk's elements
within a circle of the 'accretion' radius from the crossing points do not contribute to the
interaction between the binary and the disk.  The accretion radius is defined
as $r_{acc}=qr_c$, where $r_c$ is the radial distance of a gas element from the primary black hole.

\begin{figure}
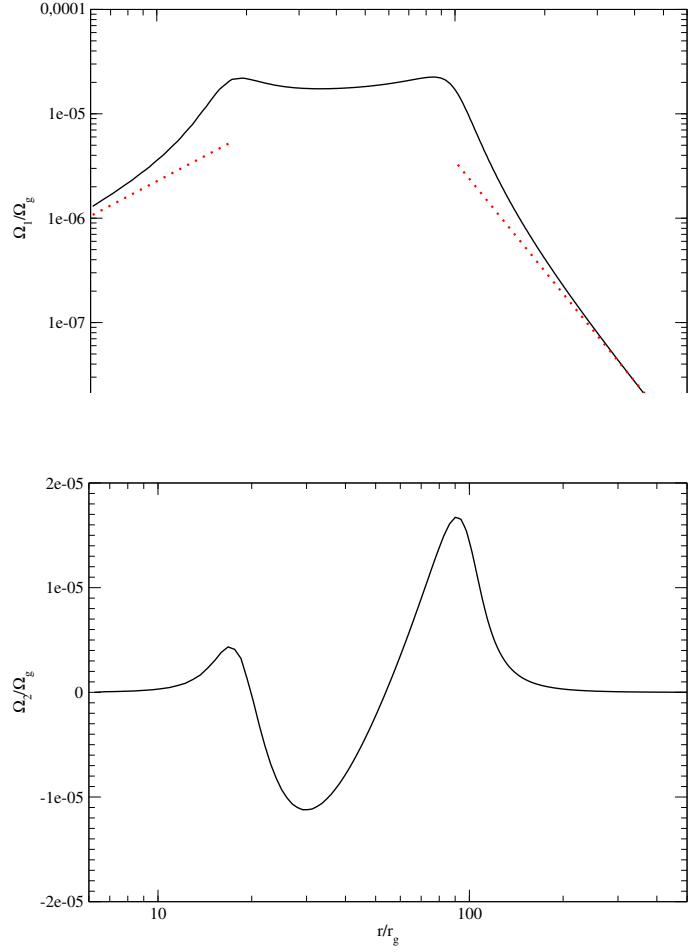

    \includegraphics[width=0.5\linewidth]{Omega1.eps}
    \includegraphics[width=0.5\linewidth]{Omega2.eps}
\caption{Top panel: The solid curve represents $\Omega_1$ expressed in units
of 'gravitational frequency' $\Omega_g=c/r_g$ as functions of $r/r_g$, respectively. The
dotted curves show the asymptotic expressions \ref{q1}. Bottom panel: The dependency of $\Omega_2$ on $r$ is shown. Unlike the corresponding expression for $\Omega_1$ it can
be negative within the orbit, when $r_p < r < r_a$.}
\label{Fig1}
\end{figure}

As shown in \cite{IZ}, under these approximations the presence of
the secondary causes the evolution of the Euler angles $\beta (t,r)$ and $\gamma (t,r)$ 
at the rate
\begin{equation}
\dot {\bf W}\equiv \dot {\bf W}_b=i\, \left [ \Omega_1(r)({\bf W}_b-{\bf W})+\Omega_2(r)e^{2i\Psi_b}(e^{2i\gamma_b}{\bf W}^{*}-{\bf W}_b)\, \right ],
\label{e17c}
\end{equation}
where $*$ stands for a complex conjugate and, in general, the functions $\Omega_1$ and
$\Omega_2$ have to be calculated numerically, see \cite{IZ} for the prescription,
${\bf W}_b=\beta_b e^{i\gamma_b}$.
Typical dependencies of  $\Omega_1$ and
$\Omega_2$ on the radial coordinate $r$  are shown in Fig. \ref{Fig1} for the binary parameters specified above as solid and dashed lines, respectively.
As dotted lines, we show approximate asymptotic analytic expressions for $\Omega_1(r)$,
which are valid when $r$ is either much smaller than $r_p$ or much larger than $r_a$
\begin{equation}
\begin{aligned}
\Omega_1\approx {3\over 4}q \left (1+{3\over 2}e^2 \right ){\left ( {a\over r} \right )}^2\Omega_k, \quad \mbox{when} \quad r \gg r_a \\
\Omega_1 \approx {3\over 4}{GqM\over \epsilon^3 a^3\Omega_k}, \quad \mbox{when} \quad r \ll r_p,
\label{q1}
\end{aligned}
\end{equation}
see \cite{IZ} and \cite{ZL}. It is important to note that, in general, both frequencies $\Omega_1$ and $\Omega_2$ depend on the angle $\Psi_b$ as well as on ${\bf W}_b$
and ${\bf W}$ at the radii corresponding to the intersections of the orbit and the disk 
\footnote{We remind that the latter dependency is caused by our requirement that the disk gas in the vicinity of the crossing points is removed from the system and does not contribute to
the averaged torque exerted by the binary on the disk.}. However, these dependencies do not change
a qualitative character of the behavior of $\Omega_1$ and $\Omega_2$ and, therefore, they
are not specified in the plot.

As was pointed out by \cite{IZ},
the main effect on the dynamics of the disk is caused by the terms
proportional to $\Omega_1$, since the terms proportional to $\Omega_2$ are largely averaged
out \footnote{See, however, the additional discussion of effects caused by $\Omega_2$ below.}.
One can see from Fig. \ref{Fig1} that $\Omega_1$ is approximately constant within the range
of radii such that $r_p < r < r_a$ and sharply falls off at the smaller and larger radii.
The value of $\Omega_1$ in the range can be estimated as $\Omega_1 \sim q\Omega_k^b$.
Comparing this estimate with the values shown in the Figure we see that the numerical factor
in this expression is close to unity and estimate hereafter that
\begin{align}
\Omega_1\approx q\Omega^b_k, \quad \mbox{when} \quad r_p < r
< r_a
\label{q2}
\end{align}

The right hand side (hereafter r.h.s.) of  Eq. \ref{e17c} is incorporated into the numerical twisted disk model as an additional term on the r.h.s. of eq. (61) of \cite{ZI}, which describes the time evolution of ${\bf W}$.
We also note that the presence of the binary leads to an additional apsidal precession of the slightly perturbed
disk's rings. The corresponding term is also included in our numerical model, but it gives only a minor
contribution to the evolution of the apsidal angle caused by the Einstein precession (see \cite{IZ}) and it not
discussed later in the text.

The quantities $\Omega_1(r)$ and $\Omega_2(r)$  determine secular interactions of the binary with the twisted disk 
and formally depend on the binary 
inclination angle $\beta_b$ due to the assumption that the disk gas within $r_{acc}$ from the binary-disk crossing points is
removed, see \cite{IZ} for the appropriate discussion. However, this dependency is not expected
to lead to any qualitative difference, and we set $\beta_b=0.1$ in this Paper, assuming, that in the framework of our model 
the ratio $\beta_{rel} =\beta/\beta_b$ remains qualitatively the same. We have checked it by considering a few test 
calculations with a larger $\beta_b=0.5$. This would allow us to make some qualitative conclusions on the applications 
of our approach to the modelling of OJ 287, where $\beta_b$ is assumed to be rather large. 

We would like to stress again, however, that our dynamical 
equations for $\beta$ and $\gamma$ may 
have unstable solutions for large values of $\beta$. In particular,
a sufficiently thin  disk may be broken into several parts as was first noted in  \cite{L1} and \cite{L2}.  
This issue is briefly analyzed in Section \ref{nonlinear} below, where we conclude that our model may be stable when
this disk is thick enough.

\subsection{Analytic treatment of the problem}

Numerical calculations performed in \cite{IZ} and in this Paper show that the twisted
disk configuration averaged over several periods of the binary's Einstein precession  tends to a stationary form in the frame precessing with the binary's Lense-Thirring frequency $\Omega_{LT}^b$ after a relatively short time interval of non-stationary
evolution, and, accordingly, we can assume that
${\bf W}\approx \hat {\bf W}(r)e^{i\Omega_{LT}^b t}$.
In order to analyze such configuration in the simplest possible way,
we use eqns (38) and (39) of \cite{DI} setting there $\dot {\bf A}=i\Omega_{LT}^b{\bf A}$ and $\dot {\bf W}=i\Omega_{LT}^b{\bf W}$ 
\footnote{We note that the auxiliary variable $\bf A$ has a simple physical meaning. It may be shown that, in the linear
approach, a twisted disk consists of ellipses with a small eccentricity. A value of the eccentricity is an odd function 
of the disk height. $\bf A$ determines this value as well as the orientation of their major axes; see \cite{II} and \cite{DI}.}.
We then use eq. (38) of \cite{DI} to express ${\bf A}$ in terms of the radial derivative of ${\bf W}$, and, substituting the result in eq. (39) of \cite{DI} and adding r.h.s. of (\ref{e17c}) to its r.h.s., we obtain
\begin{equation}
{\delta^2 \over 4} {GM\over r}{d\over dr}\left [ {\hat {\bf W}_{,r}\over (\Omega^b_{LT}-\Omega_{E}-i\alpha \Omega_{k})} \right ]=
(\Omega_{LT}-\Omega^b_{LT}-\Omega_{1})\hat {\bf W}+\Omega_{1}\beta_{b}.
\label{eq1}
\end{equation}
Note that we neglect the terms proportional to $\Omega_2(r)$ assuming that they
give a small contribution after averaging over several periods of Einstein
precession of the binary. We also stress that it is assumed in eq. (\ref{eq1}) and
all analytic estimates below that $\Omega_1(r)$ does not depend on time \footnote{\cite{IZ} used $\hat \Omega_1$ to represent the time independent
'average' value of $\Omega_1$.}.

It is convenient to represent eq. (\ref{eq1}) using a set of redefined quantities determining its coefficients.
We 
introduce
\begin{equation}
{\bf \Omega}_{c}\equiv \Omega_c e^{i\Psi}=
\alpha \Omega_{k}+i(\Omega^b_{LT}-\Omega_{E})
\label{eom}
\end{equation}
and $\Omega_s={\delta \over 2}\Omega_k$. Using these variables
eq. (\ref{eq1}) can be rewritten in a more compact form
\begin{equation}
r^2{\Omega_s^2}{d\over dr}\left [ {{\bf \Omega}_{c}^{-1}\hat {\bf W}_{,r}} \right ]=
i(\Delta \Omega \hat {\bf W}-\Omega_{1}\beta_{b}).
\label{eq1nnn}
\end{equation}
where
\begin{equation}
\Delta \Omega =\Omega_1 + \Omega^b_{LT}-\Omega_{LT}.
\label{Dom}
\end{equation}
Assuming that eq. (\ref{q2}) can adequately describe the shape of $\Omega_1$
the expression (\ref{Dom}) can be brought to a very simple form with the help
of eqns (\ref{d1}), (\ref{frecb}):
\begin{equation}
\Delta \Omega \approx q\Omega^b_K+\Omega^b_{LT}-2\chi {\left({r_g\over r}\right )^{3/2}}\Omega_K.
\label{Dom1}
\end{equation}

Note that formally the same equation can be obtained in the fully relativistic, low viscosity $\alpha \ll 1$ and slow rotation $|\chi|\ll 1$ limit from
eqns (59) and (60) of \cite{ZI} instead of eqns (38) and (39) of \cite{DI}  in the fully relativistic case, following the same
procedure. However, in this case the quantities ${\bf \Omega_c}$ and $\Omega_s$  must be redefined.

We stress that eq. (\ref{eq1nnn}) is supposed to be valid in a low viscosity $\alpha
\ll 1$, small aspect ratio $\delta < 1$ case. It is also assumed that, formally,
only radii $r \gg r_g$ are considered. Therefore, the relativistic factor $D(r)$ is
set to unity in (\ref{eq1nnn}).

Although equation (\ref{eq1nnn}) should in general be solved numerically, in the low viscosity limit specified as $\alpha < \delta$,
it has simple analytic solutions both when $r$ is formally much smaller than $r_p$ and when $r$ is much larger than $r_a$.

At first, let us consider the former case.
From Fig. \ref{Fig1} it is seen that $\Omega_1$ sharply falls off when $r$ becomes smaller than $r_p$. In the simplest approximation we can neglect the inhomogeneous term in eq. (\ref{eq1nnn}) in this region. Also, when viscosity is small
the real part of ${\bf \Omega}_c$ can be omitted \footnote{It corresponds to
the requirement that $\alpha < \chi^{-2/5}\delta^{4/5}$, see \cite{II}.}. We can also neglect $\Omega_{LT}^b$ in the imaginary part of ${\bf \Omega}_c$  at small radii, setting accordingly, ${\bf \Omega}_c=-i\Omega_{E}$.
In this case, eq. (\ref{eq1nnn}) is reduced to the simple
form describing a formally inviscid stationary twisted disk around a Kerr black hole
in the post-Newtonian approximation, see \cite{II} and also \cite{ZI}. When $\chi > 0$
it has a simple general solution in terms of Bessel functions ${\rm J}_{\nu}(y)$:
\begin{equation}
\hat {\bf W}=y^{3/5}{\hat{\bf W}_{*}}\left [ 2^{-3/5}\Gamma\left ({2\over 5}\right ) {\rm J}_{-3/5}(y)+C_{rel}{\rm J}_{3/5} (y) \right ],
\quad y={6\over 5}\sqrt{\left ({32\over 3} {\chi \over \delta^2}\right )}{\left ({r\over r_g} \right )}^{-5/4},
\label{th1}
\end{equation}
where $\Gamma (x)$ is Gamma function \footnote{We remind that eq. (37) of \cite{II} contains a misprint, the factor $y$ (denoted as $y_2$ in \cite{II}) should be
in the power $3/5$, while it has the power $3/2$ in \cite{II}.}. The solution is characterized by two constants,
$\hat{\bf W}_{*}$ and $C_{rel}$. The former constant gives the disk inclination in
the formal limit $r\rightarrow \infty$ ($y\rightarrow 0$). In our approximate treatment, it is assumed that it equals the disk inclination at
$r=r_p$. The constant $C_{rel}$ is determined by the disk behavior at small radii, where the post-Newtonian approximation implied in eq. (\ref{th1}) is not valid.
It should be determined from an asymptotic behavior of a fully relativistic
twisted disk model at $y \rightarrow 0$. In our case it is the stationary variant of the model of a twisted disk considered in \cite{ZI}.

In the opposite case $r \gg r_a$, the solution is mainly determined by
non-inertial character of the precessing frame. In this case
we can neglect all variable terms in the denominator of the
expression under the derivative on the l.h.s of eq. (\ref{eq1nnn}) and in the brackets in the front of
the first term on the r.h.s of eq. (\ref{eq1nnn}) as well as
the second inhomogeneous term on the r.h.s. of this equation. In this
way, eq. (\ref{eq1nnn}) is reduced to
\begin{equation}
{d^2\over dr^2}\hat {\bf W}={4(\Omega_{LT}^b)^2\over GM \delta^2}r\hat {\bf W},
\label{th2}
\end{equation}
which can be further reduced to the Airy equation
by the substitution
\begin{equation}
r=-r_{NI}z, \quad r_{NI}={\left ({\delta^2 GM\over 4(\Omega_{LT}^b)^2} \right )}^{1/3}\approx
8\cdot 10^2 {\left ({\delta \over \chi} \right )}^{2/3}r_g,
\label{th3}
\end{equation}
where we use the orbital parameters chosen for the binary's model to
obtain the last equality.
Accordingly, its general solution is a linear combination
of Airy functions of first and second kind, ${\rm Ai}(z)$ and ${\rm Bi}(z)$,
\begin{equation}
\hat {\bf W} = C_1 {\rm Ai} (-z)+C_2 {\rm Bi} (-z).
\label{th4}
\end{equation}
In the limit $|z| \gg 1$, we obtain
\begin{equation}
\hat {\bf W} \approx {1\over \sqrt \pi |z|^{1/4}} \left [ C_1 \sin \left (
{2\over 3}|z|^{3/2}+{\pi \over 4} \right )+
C_2 \cos \left ( {2\over 3}|z|^{3/2}+{\pi \over 4} \right ) \right ].
\label{th4a}
\end{equation}
Remembering that ${\bf W}=\hat {\bf W}e^{i\Omega_{LT}^bt}$, we see from (\ref{th4a}) that the solution far from the binary's orbit is a linear combination of ingoing
and outgoing waves proportional to $e^{i(\Omega_{LT}^b t\pm {2\over 3}|z|^{3/2})}$,
where plus (minus) corresponds to ingoing (outgoing) solution. Since we assume
that there is no influence on the disk's dynamics from the side of large radii,
the part of the solution corresponding to an ingoing wave should be set to zero. This
determines the outer boundary condition for our problem. It is easy to see from
(\ref{th4a}) that this condition takes the form $C_1=-iC_2\equiv -iC$, and we finally
have
\begin{equation}
\hat W = C({\rm Bi} (-z)-i {\rm Ai}(-z)).
\label{th5}
\end{equation}


\section{Approximate analytic analysis of the twisted disk configurations in
the low viscosity case}
\label{analyt}

When analyzing solutions of eq. (\ref{eq1nnn}), 
it is important to point out
that they 
are qualitatively different in the cases of very thin disks
with $\delta \ll 1$ and relatively thick disks with $\delta \lesssim 1$. A concrete criterion delineating these cases is specified below.
It turns out
that a disk should have its relative thickness $\delta \gtrsim 0.1$
to be considered as a relatively thick disk, see derivation of eq. (\ref{ne17n}) below.  We consider these cases in their turn in the subsequent Sections.

\subsection{The case of a very thin disk $\delta \ll 1$: generation of waves of tilt and twist near the resonance}

When a disk with $\delta \ll 1$ is considered, similar to the case of the large
viscosity $\alpha=0.1$ considered in \cite{IZ}, an important role is played
by the elementary 'equilibrium' solution, $\hat {\bf W}\approx \hat {\bf W}_{eq}$, which is obtained from (\ref{eq1nnn})
when the l.h.s proportional to $\delta^2$ is neglected and, accordingly,
\begin{equation}
\hat {\bf W}_{eq}\equiv \beta_{eq}={\Omega_{1}\beta_{b}\over \Delta \Omega}.
\label{equ1}
\end{equation}
Clearly, this solution is invalid close to the radius $r_r$, where
\begin{equation}
\Delta \Omega (r_r)=\Omega_{LT}(r_r)-\Omega^b_{LT}-\Omega_1(r_r)=0.
\label{equ2}
\end{equation}
Physically, this is a resonant condition between the Lense-Thirring frequency of
a particular disk ring calculated in the precessing frame, $\Omega_{LT}(r_r)-\Omega^b_{LT}$ and the forcing frequency
$\Omega_1$ determining the influence of the secondary on the disk.

With the help of eq. (\ref{Dom1}) we can obtain a useful approximate expression for the location of the resonance
\begin{equation}
r_r\approx 33{\left ( {2.5\chi_{*}\over (1.5q_{*}+\chi_{*})} \right ) }^{1/3}r_g,
\label{rr}
\end{equation}
where we introduce $\chi_{*}=\chi/0.5$, $q_*=q/7.5\cdot 10^{-3}$, and we set $a=60r_g$ and $\Omega_{LT}^{b}\approx 10^{-2}\chi \Omega_{K}^b$ corresponding to the binary
orbital parameters used in this study.

In \cite{IZ} it was shown that when viscosity is large, one can find an approximate solution
of eq. (\ref{eq1nnn}) in the vicinity of the resonance by decomposing the coefficients entering this equation in Taylor series in $(r-r_r)/r_r$ which is expressed in terms of the so-called Scorer functions 
\footnote{We remind that these functions are solutions of an inhomogeneous Airy
equation.}. This solution can be smoothly matched to the equilibrium solution
(\ref{equ1}). It was also shown that this analytic result is in good agreement with the numerical simulations.

When $\alpha$ is small enough, numerical results suggest that there is another contribution to $\hat {\bf W}$, which has a typical characteristic radial scale of variation
much smaller than $r$. It is natural to assume that, away from the resonance, this
contribution is a WKBJ wave (possibly overdamped), which is launched at the
resonance in a way similar to the well-known mechanism of excitation of density waves
at Lindblad resonances, see e.g. \cite{GT}. Accordingly, we represent $\hat {\bf W}$
as
\begin{equation}
\hat {\bf W} =\hat {\bf W}_{eq}+\hat {\bf W}_{wave}.
\label{equ3}
\end{equation}
In order to find $\hat {\bf W}_{wave}$, we consider again solutions of eq. (\ref{eq1nnn})
in the region of small $|(r-r_r)|/r_r$. Using the solution in
terms of Scorer functions found in \cite{IZ}, we note that there is an additional
contribution to its asymptotic form at large values of its argument (but still
within the region  $|(r-r_r)|/r_r \ll 1$), which can be interpreted as a $\hat {\bf W}_{wave}$. This contribution was neglected in \cite{IZ}, since its absolute value
is small when $\alpha$ is sufficiently large, but it plays a role of
the source of the wave-like contribution to $\hat{\bf W}$ at smaller values of
$\alpha$. It is considered below, see Section \ref{res}.
We then consider a general WKBJ solution of eq. (\ref{eq1nnn}), see Section \ref{wkbj}, and match it to the found asymptotic solution, thus obtaining a wave-like
solution valid for all radii. A comparison of this approach with the numerical
results is made in Section \ref{numerical}.

\subsubsection{An approximate solution of eq. (\ref{eq1nnn}) near the resonance}
\label{res}

Let us consider a region close to $r_r$, where $x=(r-r_r)/r_r \ll 1$. In this region we can assume that ${\bf \Omega}_c$, $\Omega_s$ and $\Omega_1$ are equal to their values at $r_r$,
and approximate $\Delta \Omega$ as
\begin{equation}
\Delta \Omega =\Omega_{res} x,
\label{ne2n}
\end{equation}
where $\Omega_{res}$ does not depend on $r$. In what follows, we assume that $\Omega_{res}$ is positive, which is valid
when $\chi > 0$. We note, however, that in the opposite case $\chi < 0 $,
our treatment remains the same provided we change $x\rightarrow -x$ 
\footnote{\cite{IZ} used $\Omega_{*}$ instead
of $\Omega_{res}$. Note a misprint in their eq. (58), where the sign minus should be absent in the front
of $\Omega_{*}$. This misprint does not propagate further in the text.}.
We can use the expressions (\ref{Dom1}) and (\ref{rr}) to obtain an approximate
estimate
\begin{equation}
\Omega_{res}\approx \tilde \Omega_{res}\Omega^{(r)}_{K}, \quad \mbox{where} \quad \tilde \Omega_{res}\approx 1.6\cdot 10^{-2} \chi_{*} { \left ( {33r_g\over r_r} \right ) }^{3/2}.
\label{Omres}
\end{equation}
Hereafter the superscript $(r)$ denotes all frequencies of interest
evaluated at $r=r_r$, e.g. ${\bf \Omega}^{(r)}_c={\bf \Omega}_c(r=r_r)$, $\Omega^{(r)}_s=\Omega_s(r=r_r)$, $\Omega^{(r)}_1=\Omega_1(r=r_r)$, $\Omega_K^{(r)}=\Omega_K(r=r_r)$, $\Omega_{E}^{(r)}=\Omega_{E}(r=r_r)$ and $\Omega_{LT}^{(r)}=\Omega_{LT}(r=r_r)$.
The expression (\ref{Omres}) turns out to be
in a good agreement with the results of numerical calculations.

After the assumption above have been made, eq. (\ref{eq1nnn}) can be brought to the form equivalent to that
of eq. (60) of \cite{IZ}
\begin{equation}
{d^2\over dx^2}\hat{\bf W}=i({\mbox{\boldmath$\omega$}}_{res} x \hat{\bf W} -{\mbox{\boldmath$\omega$}}_{1} \beta_b),
\label{ne3n}
\end{equation}
where
\begin{equation}
{\mbox{\boldmath$\omega$}}_{res}={\Omega_{res}{\bf \Omega}^{(r)}_{c}\over {\Omega^{(r)}_{s}}^2},
\quad {\mbox{\boldmath$\omega$}}_{1}={\Omega^{(r)}_{1}{\bf \Omega}^{(r)}_{c}\over
{\Omega^{(r)}_{s}}^2}.
\label{ne4n}
\end{equation}

However, unlike the case considered in \cite{IZ},
now both $\omega_{r}$ and $\omega_{1}$ are, in general, complex
quantities. They become real when the imaginary part of  eq. (\ref{eom}) is neglected,
and, accordingly, $\Psi=0$.

Similar to \cite{IZ},
the solution to eq. (\ref{ne3n}) having the property that its amplitude decays when $|x| \rightarrow \infty $ can be expressed in terms of the Scorer function, ${\rm Hi(}z)$
\begin{equation}
\hat{\bf W}=i\pi {{\mbox{\boldmath$\omega$}}_{1}\over {\mbox{\boldmath$\omega$}}^{2/3}_{res}}\beta_b {\rm Hi}({\bf z}), \quad \mbox{where} \quad {\bf z}=-i{\mbox{\boldmath$\omega$}}_{res}^{1/3}x=
-i{\Omega_{res}^{1/3}{{\bf \Omega}^{(r)}_{c}}^{1/3}\over {\Omega^{(r)}_{s}}^{2/3}}x.
\label{ne5n}
\end{equation}
As discussed in e.g. \cite{Olver}, when $|z|$ is large the Scorer function has the following asymptotic
form
\begin{equation}
{\rm Hi}({\bf z})\approx -{1\over \pi {\bf z}}+{1\over \sqrt{\pi}{\bf z}^{1/4}}\exp {\mbox{\boldmath$\zeta$}},
\quad \mbox{where} \quad {\mbox{\boldmath$\zeta$}}={2\over 3}{\bf z}^{3/2}.
\label{ne6n}
\end{equation}
Note that the first term describes the stationary solution (\ref{equ1})
in the vicinity of the resonance and also note that the last term was discarded in \cite{IZ}.

In general, eq. (\ref{ne6n}) provides a valid solution only when the last term decays
when $|{\mbox{\boldmath$\zeta$}}|$ grows. This condition requires $\Re[{\mbox{\boldmath$\zeta$}}]$ to be negative. To show that
this always holds for the problem at hand, we represent $x$ as $e^{i\nu}|x|$, where
$\nu =0 $ when $x > 0$ and $\nu =\pi$ otherwise. Now, using eqns (\ref{eom}), (\ref{ne4n}), (\ref{ne5n})
and (\ref{ne6n}) we represent ${\mbox{\boldmath$\zeta$}}$ as
\begin{equation}
{\mbox{\boldmath$\zeta$}}={2\over 3}\omega_{res}^{1/2}|x|^{3/2}\exp \left [ i \left ( {3\over 4}(2\nu - \pi)+{\Psi\over 2} \right ) \right ],
\label{ne7n}
\end{equation}
where $\omega_{res}=|{\mbox{\boldmath$\omega$}}_{res}|$
Substituting $\nu=0$ and $\pi$ in eq. (\ref{ne7n}) we obtain
\begin{equation}
{\mbox{\boldmath$\zeta$}}={2\over 3}\omega_{res}^{1/2}|x|^{3/2} \left [ \cos \left ({3\pi\over 4}\mp {\Psi\over 2} \right )\mp i\sin \left ( {3\pi\over 4}\mp {\Psi\over 2} \right ) \right ],
\label{ne8n}
\end{equation}
where $(-)$ and $(+)$ correspond to positive and negative values of $x$, respectively.
Noting that from the definition of $\Psi$ by eq. (\ref{eom}), it follows that as soon as $\alpha $ is positive,
it is enough to consider its values in the range $-{\pi\over 2} < \Psi < {\pi \over 2}$.
Therefore, we see from eq. (\ref{ne8n}) that $\Re[{\mbox{\boldmath$\zeta$}}]$ is always negative for these values of $\Psi$
at either side of the resonance.
However, when the sign in eq. (\ref{ne8n}) is positive and $\Psi \rightarrow -{\pi\over 2}$ or it's negative and
 $\Psi \rightarrow {\pi\over 2}$ the real part of $\Re[{\mbox{\boldmath$\zeta$}}]$ tends to zero. In these
cases, in addition to the stationary solution (\ref{equ1}), we have an almost undamped wave launched
outward from the
resonance when $\Omega^{b}_{LT} > \Omega^{(r)}_{E}$ and launched inward in the opposite case \footnote{Note
that when $\Omega_{res} < 0$ this condition is reversed.}. $|\Psi|$ is close to ${\pi\over 2}$
when
\begin{equation}
\alpha \ll {|\Omega^{b}_{LT} - \Omega^{(r)}_{E}|\over \Omega^{(r)}_{K}}.
\label{ne9n}
\end{equation}

\subsubsection{A WKBJ analysis of the wave-like contribution}
\label{wkbj}

In the above discussion of the wave-like contribution to the solution of eq. (\ref{eq1nnn}),
it was assumed that $|x|=|r-r_r|/r_r$ is small. At larger scales,
it can be matched to a WKBJ solution of eq. (\ref{eq1nnn}), which describes free stationary (in the precessing frame) waves with wavelengths much shorter than $r$. These waves
are overdamped in case when $|\Psi|$ isn't close to $\pi/2$. In order to find such solutions, we set the source term proportional to
$\beta_b$ in eq. (\ref{eq1nnn}) to zero and make the standard WKBJ anzatz $\hat{\bf W}_{wave}=\hat{\bf A}(r)\exp \left ( {i\int_{r_r}^{r}dr^{\prime}{\bf k}(r^\prime)} \right )$.
Substituting this expression into eq. (\ref{eq1nnn}) and neglecting the terms that are proportional to the second derivative
of $\hat{\bf A}$
, we find
\begin{equation}
\hat {\bf W}_{wave}={r^{1/2}{\bf \Omega_c}^{1/2}\over {\Delta \mbox{\boldmath$\omega$}}^{1/4}}
\left ( {\bf C_1}e^{i{\bf \phi}_{WKBJ}}+{\bf C_2}e^{-i{\bf \phi}_{WKBJ}} \right ),
\label{ne10n}
\end{equation}
where ${\bf C}_{1,2}$ are constants,
\begin{equation}
 \Delta {\mbox{\boldmath$\omega$}}={{\bf \Omega}_{c}\Delta \Omega \over \Omega_s^2},
\end{equation}
and
\begin{equation}
{\bf \phi}_{WKBJ}=e^{3\pi i/4}\int_{r_r}^{r}{dr^{'}\over r^{'}}\sqrt{\Delta {\mbox{\boldmath$\omega$}}}.
\label{ne11n}
\end{equation}

Comparing eq. (\ref{ne10n}) taken in the limit $r\rightarrow r_r$ with the second term in eq. (\ref{ne6n}), we find that they agree
with each other provided
$$
{\bf C}_1=\sqrt \pi e^{i({5\pi/8}+\Psi/6)} {\Omega^{(r)}_{1}\over r_{r}^{1/2} \Omega^{(r)}_{s}{\Omega_{res}}^{1/2}}, \quad {\bf C}_2=0,
$$
where we remind that the upper index $(r)$ implies
that the corresponding quantities are evaluated at the resonance position $r=r_r$. Thus, in general, when $|z|$ defined in eq. (\ref{ne5n}) is
large, the solution of eq. (\ref{eq1nnn}) has the form of (\ref{equ3}),
where
\begin{equation}
\hat {\bf W}_{wave}=\sqrt{\pi}{\Omega^{(r)}_1{\bf \Omega}_c^{1/2}\over \Omega^{(r)}_{s}{\Omega_{res}}^{1/2} {\Delta \mbox{\boldmath$\omega$}}^{1/4}}
{\left ({r\over r_r}\right )}^{1/2} \beta_b e^{i({\bf \phi}_{WKBJ}+5\pi/8+\Psi/6)}.
\label{ne13n}
\end{equation}
From eq. (\ref{ne13n}) it follows that
\begin{equation}
\beta_{wave}\equiv |\hat{\bf W}_{wave}|=\sqrt{\pi}{\Omega^{(r)}_1{\Omega}_c^{1/2}\over \Omega^{(r)}_{s}{\Omega_{res}}^{1/2} {|\Delta\mbox{\boldmath$\omega$}|}^{1/4}}
{\left ({r\over r_r}\right )}^{1/2}\beta_b\, \exp \left (-\int^{r}_{r_r}{dr^{'}\over r^{'}}|{\Delta \mbox{\boldmath$\omega$}}|^{1/2}\sin \left( {\Psi\over 2}+{3\pi\over 4}+{\nu\over
2}  \right)\, \right),
\label{ne14n}
\end{equation}
where we remind that $\nu=0$ when $r > r_r$ and $\nu=\pi$ otherwise.
 Recalling that $\Psi$ is defined such that
$-{\pi\over 2} < \Psi < {\pi\over 2}$ we see from eq. (\ref{ne14n}) that the integrand is positive when $r > r_r$
and it is negative otherwise. This ensures that the expression under the exponent is always negative.
It is also equal to zero when $\Psi =\pi/2$ and $r > r_r$ or when $\Psi=-\pi/2$ and $r < r_r$. Both these
limiting cases formally correspond to $\alpha =0$. In the former case, we also have $\Omega_{LT}^{b} > \Omega^{(r)}_{E}$,
while the latter corresponds to $\Omega_{LT}^{b} < \Omega^{(r)}_{E}$. As seen in the precessing frame, these cases describe the launching
of a non-dissipative stationary wave at the resonance, while the difference
$\Omega_{LT}^{b} - \Omega^{(r)}_{E}$ determines the direction of its propagation.

In order to find an explicit value of $\beta_{wave}$, one should be able to evaluate
numerically the profile of $\Omega_1(r)$. However, when $\alpha$ is small and only
the region within the orbit, $r_p < r < r_a$, is considered, there is a possibility
to obtain an analytic approximate expression for $\beta_{wave}$, which agrees with
the numerical one up to the factor $\sim 1.5-2$, see Fig. \ref{Fig5} below. This can be done as follows.

At first, we can rewrite the expression (\ref{Dom1}) in an alternative form using
eqns (\ref{q2}), (\ref{rr}) and (\ref{Omres}). We get
\begin{equation}
\Delta \Omega = {\tilde \Omega_{res}\over 3}\,(1-{\tilde r}^{-3})\,\Omega_{K}^{(r)}, \quad \mbox{where}
\quad \tilde r ={r\over r_r}
\label{Dom2}
\end{equation}
Secondly, in the limit of small $\alpha$,
the leading term in the expression for $\Omega_c$ is the Einstein apsidal
precession frequency, $\Omega_{E}={3GM\over c^2r}\Omega_K=3{\tilde r}^{-5/2}\Omega_{K}^{(r)}$.
Accordingly, we have
\begin{equation}
{\bf \Omega}_c\approx -i\Omega_E\, \left (1+i{\alpha \Omega_K\over \Omega_E} \right ),
\label{beta1}
\end{equation}
where it is assumed that the second term in brackets is much smaller than unity. 
Assuming that $r < r_r$, and setting, therefore,
$\nu=\pi$
in eq. (\ref{ne14n})
we use eq. (\ref{beta1}) to obtain
$\Psi \approx -{\pi \over 2}+{\alpha \Omega_k\over \Omega_{E}}$ and further estimating
\begin{equation} \sin \left ( {\Psi \over 2}+{3\pi\over 4} +{\nu\over 2} \right )\approx -{\alpha \Omega_k\over 2\Omega_{E}}=-{\alpha r_r \tilde r\over 6r_g}.
\label{sin}
\end{equation}
We note that in the opposite case of $r > r_r$ and, accordingly,  $\nu=0$,
$\sin ({\Psi \over 2}+{3\pi\over 4} +{\nu\over 2})\approx 1$, and, therefore,
the wave-like part of the twist exponentially decays with $r$.

In the same approximation, we have
\begin{equation}
|\Delta {\mbox{\boldmath$\omega$}}|\approx {4\over \delta^{2}}{r_g\over r_{r}}\tilde \Omega_{res}
{\tilde r}^{-5/2}|1-{\tilde r}^{3}|.
\label{nbeta2}
\end{equation}

We represent eq. (\ref{ne14n}) as $\beta_{wave}=\hat A\exp{(-\sigma )}$ and evaluate the
factor $\hat A$ using eqns (\ref{q2}), (\ref{Dom2}) and (\ref{sin}):
\begin{equation}
\hat A\approx \sqrt{6\pi}\, {\left ( {r_g\over r_r} \right )}^{1/4}q\beta_b\, \delta^{-1/2}{\tilde \Omega_{res}}^{-3/4} \, {{\tilde r}^{-1/8}\over (1-{\tilde r}^{3})^{1/4}}.
\label{nbeta3}
\end{equation}
The factor $\sigma $ can be evaluated with the help of eqns (\ref{sin}) and (\ref{nbeta2})
\begin{equation}
\sigma \equiv \int^{r}_{r_r}{dr^{'}\over r^{'}}|{\Delta \mbox{\boldmath$\omega$}}|^{1/2}\sin \left ( {\Psi\over 2}+{3\pi\over 4}+{\nu\over
2}\right )={\alpha \over 3\delta }{\left ({\tilde \Omega_{res} r_r\over r_g} \right )}^{1/2}I, \quad \mbox{where} \quad I=\int^{1}_{\tilde r}
dx x^{-5/4}(1-x)^{1/2}.
\label{phi}
\end{equation}
The integral $I$ in eq. (\ref{phi}) can be expressed in terms of the incomplete beta function
${\rm B}(a;b;x)\equiv \int_0^x dy\, y^{a-1}(1-y)^{b-1}$ as $I=4{\tilde r}^{-1/4}(1-{\tilde r}^3)^{1/2}-2{\rm B}({1\over 2};{11\over 12};
1-{\tilde r}^3)$.

Since the expression (\ref{nbeta3}) is obtained in WKBJ approximation,
it is valid only
when] $|z|$ defined in eq. (\ref{ne5n}) is larger than unity. This condition yields:
\begin{equation}
\left | {(r-r_r)\over r_r} \right | > \left ( {\delta\over 2} \right )^{2/3}\left ({r_r \over 3r_g} \right )^{1/3}\,{\tilde \Omega_{res}}^{-1/3}.
\label{nbeta4}
\end{equation}
On the other hand, the l.h.s. of eq. (\ref{nbeta4})
should be smaller than unity. The latter condition results in
\begin{equation}
\delta < 2\,{\left ( {3r_g\over r_r} \right )}^{1/2}\,{\tilde \Omega_{res}}^{1/2}.
\label{ne16n}
\end{equation}
Taking into account that for the model we consider in this Paper $r_r\approx 30\,r_g$
and $\Omega_r\approx 1.6\cdot 10^{-2}\,\Omega_{K}^{(r)}$, we conclude that
the twisted disk configuration is expected to be quite different
from that described in this Section provided
\begin{equation}
\delta \gtrsim 0.1.
\label{ne17n}
\end{equation}

From equation (\ref{ne6n}), it is evident that when $|z| \sim 1$ both equilibrium and wave-like
contributions to $\hat{\bf W}$ are of the same order. In the situation when the factor $\sigma$ is small, $\beta_{wave} \approx \hat A$, and from eq. (\ref{nbeta3})
it follows that $\hat A$ decreases slower than $\beta_{eq}$ with decreasing $\tilde r$.
Thus, in this case
the wave-like contribution to the solution is more important than the equilibrium one
when $\tilde r < 1$.
The condition $\sigma <1$
results in
\begin{equation}
\alpha < 3\, {\left ({\tilde \Omega_{res} r_r\over r_g} \right )}^{-11/2}I^{-1}.
\label{ne18n}
\end{equation}
Using our 'typical' values $\tilde \Omega_{res}=1.6\cdot 10^{-2}$ and
$r_r/r_g \approx 30$ and assuming $I=0.5$, which corresponds to $\tilde r\approx 0.5$,
we obtain
\begin{equation}
\alpha \lesssim 10\, \delta
\label{ne19n}
\end{equation}
When the condition (\ref{ne19n}) is not satisfied, the wave-like contribution is not important. Such a situation
was considered in our previous paper \cite{IZ}.

\subsection{An estimate of the inclination of a relatively thick twisted disk
within the orbit}
\label{thick}

Let us consider the situation when condition (\ref{ne17n}) is satisfied, and, accordingly, the
analysis of the previous Section is not applicable. Numerical results suggest that
in this case the disk also tends to align with the precessing plane.
However, the
term on the l.h.s.
of eq. (\ref{eq1nnn}) is now important at all radii, and this property complicates the analytical analysis of the situation. However, a qualitative description
is still possible taking into account that various simplifications can be made at sufficiently small, intermediate and large radii.
Accordingly, we divide the whole available range of $r$ onto three regions: 1) the inner region
$6 < r < r_p\approx 18r_g$, 2) the intermediate region $r_p < r < r_a=102r_g$, and
3) the outer region $r > r_a$. We are mostly interested in characteristic values
of the disk inclination $\beta$ in the intermediate region, since a number of
disk's crossings by the secondary is determined by a typical value of inclination
in this region. However, in order to estimate it, we need to consider an approximate
behavior of inclination in all three regions and match them together.

From Fig. \ref{Fig1} it is seen that $\Omega_1$ drops sharply when $r$ becomes smaller than $r_p$ and larger than $r_a$.
Accordingly, we assume that the inhomogeneous term in (\ref{eq1nnn}) is small in both regions, and the solutions
are approximately given by eqns (\ref{th1}) and (\ref{th5}) in the
case when $r < r_p$ and $r > r_a$, respectively. Since both equations are  homogeneous, they  do not fix the amplitude of inclinations.
However, we are going to evaluate the ratios (i) $f_{p}=-\beta^{'}(r_p)/\beta (r_p)$ and (ii) $f_{a}=-\beta^{'}(r_a)/\beta (r_a)$ with the help of these equations
and use these ratios as boundary conditions for the approximate solution of eq. (\ref{eq1nnn}) in the intermediate region $r_p < r < r_a$.

\begin{itemize}
\item[i)]
The ratio $f_p$.
Substituting $r=r_p=18r_g$ in the expression for $y$ given in eq. (\ref{th1})
we see that it is smaller than unity for the expected parameters of the thick
disk,

$$y(r_p)\approx 0.7 \left ( {0.1\over \delta} \right ) { \left ( {\chi \over 0.5} \right ) }^{1/2}.$$

We can therefore use the known asymptotic expressions for Bessel functions in
the formal limit $y\rightarrow 0$ to estimate $f_{p}$,
\begin{equation}
\hat {\bf W}\approx {\hat {\bf W}}_{*} \left [ 1 + B_{rel}{\chi}^{3/5}{\delta}^{-6/5}{\left ({r\over r_g} \right )}^{-3/2} \right ], \quad \mbox{where} \quad B_{rel}={2^{27/10}\over 5^{1/5}3^{2/5}\Gamma (3/5)}\, C_{rel},
\end{equation}
which finally gives

\begin{equation}
f_p\approx {3\over 2}\,  {B_{rel}\chi^{3/5}\delta^{-6/5}{r_g}^{3/2}r_{p}^{-5/2}\over {1+B_{rel}
\chi^{3/5}\delta^{-6/5}{(r_g/r_{p})}^{3/2}}}.
\label{fat1}
\end{equation}
As we have mentioned  after eq. (\ref{th1}), the constant $B_{rel}$ (or $C_{rel}$) cannot be obtained
in a post-Newtonian approach to the dynamics of twisted disks. A fully relativistic
treatment is required for its determination. For definiteness, we numerically obtain
this constant using the model proposed in \cite{ZI}. The result is shown in Fig. \ref{Fig2} for $\chi=0.25$ and $0.5$ and a range of rather large values of
$\delta$ \footnote{We note that this constant is expected to be model dependent.
However, we believe that using a different relativistic model won't change
our qualitative conclusions.}.
\begin{figure}
    \includegraphics[width=0.5\linewidth]{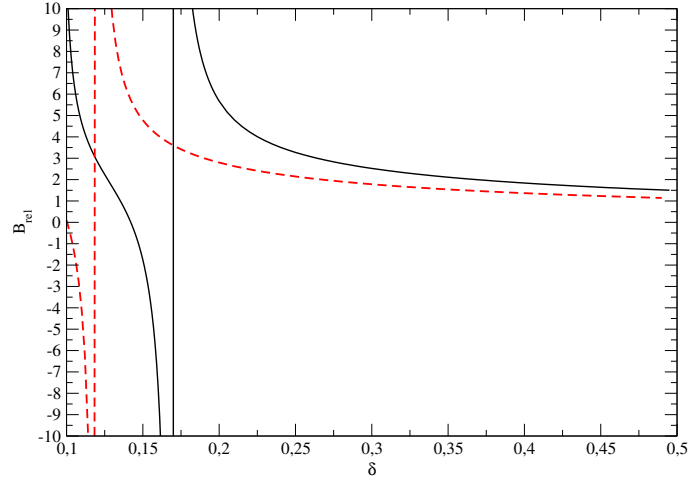}
\caption{
The dependency of $B_{rel}$ on $\delta$ is
shown. Solid and dashed curves correspond to $\chi=0.5$ and $0.25$, respectively. 
}
\label{Fig2}
\end{figure}
As shown in this Fig., $B_{rel}$ is not monotonic. It can take negative values
when $\delta$ is sufficiently small, see \cite{ZI} for an analytic calculation
of this quantity in the limit of small $\delta$. In this work we use
only two 'large' values of $\delta=0.2$ and $0.5$ and one 'intermediate'
$\delta=0.05$. The latter value is used to find an approximate boundary between 'large' and
'small' values of $\delta$ and that the value of $B_{rel}(\delta=0.05)$ is negative is not important
for this purpose.
 In the case of 'large' $\delta$, we have
$B_{rel}(\delta=0.2)\approx 5.7$ and $B_{rel}(\delta=0.5)\approx 1.5$ when $\chi=0.5$.
\item[ii)]
The ratio $f_a$.
In order to calculate $f_a=-\beta^{'}(r_a)/\beta (r_a)$, we use the solution (\ref{th5}), which is approximately
valid at large radii.
As seen from eq. (\ref{th3}), $r_{NI} > r_{a}$ when $\delta > 0.1$.
This implies that when estimating $f_{a}$, we can use the well known asymptotic expressions of Airy functions
and their derivatives valid in the limit $|z|\rightarrow 0$. After a straightforward
calculation we obtain
\begin{equation}
f_a\approx \left ( {3^{1/3}\over 2} \right ) \left ( {\Gamma (2/3)\over \Gamma (1/3)} \right ){1\over r_{NI}}
\approx {0.36\over r_{NI}}.
\label{fat3}
\end{equation}

\end{itemize}

In order to find the solution in the region $r_p < r < r_a$, we neglect the real part of ${\bf \Omega}_c$, see eq. (\ref{eom}), assuming that viscosity
is sufficiently small. In this case, eq. (\ref{eq1nnn}) becomes real and we can
set $\hat {\bf W}=\beta(r)$. We also use the expression (\ref{q2}) for $\Omega_1$.

Contrary to the previous Section, the term proportional to $\hat {\bf W}$ on r.h.s.
of eq. (\ref{eq1nnn}) doesn't play a significant role, and we neglect it
In this case, this equation becomes elementary
\begin{equation}
{d\over d \hat r}\left [ {1\over \left (\hat \Omega_{LT}^{b}-3 \hat r_g\,{\hat r}^{-5/2}\right )} \, {d\beta\over
d \hat r} \right ] ={4q\beta_b \hat r\over \delta^2},
\label{fat4}
\end{equation}
where we introduce the new dimensionless radial variables $\hat r =r/a$ and $\hat r_g = r_g/a$. Also
$$\hat \Omega_{LT}^b=\Omega_{LT}^b/\Omega_K^b={2\chi \over \epsilon^3}\,{\hat r_g}^{3/2}.$$

It is not difficult to integrate eq. (\ref{fat4}) with the result
\begin{equation}
\beta={2q \beta_b \over \delta^2} \left ({\hat \Omega_{LT}^b {\hat r}^3\over 3}-6\hat r_g\, {\hat r}^{1/2} \right )+C_1 \left ( \hat \Omega_{LT}^b {\hat r} +2\hat r_g\,{\hat r}^{-3/2} \right )+C_2.
\label{fat5}
\end{equation}
The integration constants $C_{1,2}$ should be chosen in such a way that
the conditions (\ref{fat1}) and (\ref{fat3}) are satisfied at $r=r_p$ and
$r=r_a$, respectively. This results in two linear algebraic equations
for $C_{1,2}$, which are readily solved.
We compare this solution with the results of numerical
simulations below, see Fig. \ref{Fig7} in Section \ref{numerical}. It is seen
that typical analytic values of $\beta$ in the region $r_p < r < r_a$ are larger
than the numerical values by a factor of $1.5-2$, which appears to be satisfactory,
taking into account our oversimplified analytic treatment. It is also seen from
these Figs that both analytic and numerical values of $\beta$ do not vary
much in this region. Therefore, we can use a particular value of $\hat r$ in
(\ref{fat5}) to estimate a typical value of inclination within the orbit.

As seen from Figs. \ref{Fig6} and \ref{Fig7} below, when the disk is relatively thick,
the inclination, $\beta$, is nearly constant in the region $r_p < r < r_a$. In this case,
one can simplify eq. (\ref{fat5}), assuming that only the constant $\beta_{const}\equiv C_2$
gives the main contribution to the expression. In order to estimate its value, we note
that  the inhomogeneous term in eq. (\ref{fat5}) gives the main contribution to the radial
derivative of the expression at $\hat r \sim r_a/a=(1+e)$. We formally set, accordingly, $C_1=C_2=0$
in eq. (\ref{fat5}), differentiate the expression over $\hat r$ and evaluate it at $\hat r=(1+e)$.
We then obtain $\beta_{const}\approx -{d\beta \over d \hat r}(\hat r=(1+e))/af_{a}$, where
$f_a$ is given by eq. (\ref{fat3}). We then substitute the values of $a$ and $e$ corresponding
to our model in the resulting expression to obtain
\begin{equation}
\beta \approx \beta_{0} \approx 2.8 \chi^{-2/3}\delta^{-4/3}\, (1-0.7\chi)\, q \beta_b.
\label{fat6}
\end{equation}
From eq. (\ref{fat6}) it follows that, in order to have $\beta < \beta_b$ within the orbit,
and, accordingly, only two crossings of the disk by the secondary per one orbital period,
we should approximately require
\begin{equation}
\delta \gtrsim 0.1 {\left ({0.5\over \chi} \right )}^{1/2} {\left ({q\over 10^{-2}} \right )}^{3/4}(1-0.7\chi)^{3/4}.
\label{fat7}
\end{equation}
We compare below eqns (\ref{fat6}) and (\ref{fat7}) with the results of numerical calculations
in Fig. \ref{Fig7}.

\section{Numerical calculations and the comparison to the analytic approach}
\label{numerical}

In our numerical work, we consider three values of the disk relative thickness. A thin disk is represented by $\delta=2\cdot 10^{-3}$,
while a thick disk is represented by $\delta=0.2$ and $0.5$. In addition, we consider the case
with $\delta=0.05$  as an illustration of a disk with an 'intermediate' relative thickness.
The rotational parameter is always considered to be positive in this study, $\chi > 0$,
but it is expected that our main conclusions are unchanged for the retrograde black hole rotation
with $\chi < 0$. For definiteness, we use  $\chi=0.5$ in this study.

We consider two values of $\alpha$, $\alpha=2\cdot 10^{-2}$ and $2\cdot 10^{-4}$ for the thin disk to consider  the disks $\alpha > \delta$ and vice versa. The effect of a change of $\alpha$ on the disk's dynamics appears to be rather insignificant when $\alpha$ is much smaller than $\delta$ and we set $\alpha=0.02$ for all thicker disk models.

\begin{figure}
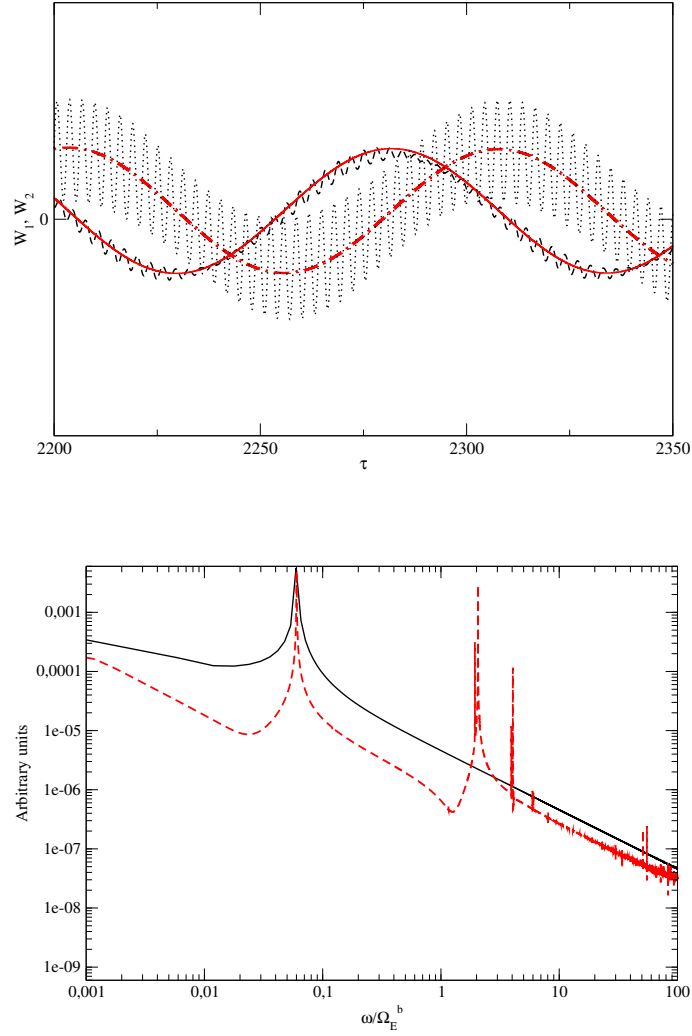

    \includegraphics[width=0.5\linewidth]{W12.eps}

\vspace{1.2cm}

\includegraphics[width=0.5\linewidth]{spectr.eps}
\caption{
Top panel.
We show the time dependencies of $W_1$ and $W_2$ as functions of time $\tau$  at $r=40r_g$. The dashed and the dotted (black)
curves describe, respectively, $W_1$ and $W_2$ obtained in the $(dn)$ model. 
The solid and dot-dashed (red) curves describe
$W_1$ and $W_2$, respectively, in the $(st)$ model. Bottom panel. The spectra of $W_2$
are shown as functions of the ratios of the frequency $\omega$ to $\Omega_E^b$. Solid (black) and dashed (red)
curves correspond to the $(st)$ and $(dn)$ models, respectively. Both models have well pronounced
peak at  $\omega/\Omega_E^b \approx 0.06$.
This ratio is close to  $\Omega_{LT}^b/\Omega_E^b$
calculated for the used $\chi=0.5$. The $(dn)$ model has, in addition, a well pronounced peak
at $\omega/\Omega_E^b \approx 2$ as well as a number of overtones.}
\label{Fig3}
\end{figure}

As explained above, the influence of binary on the disk is modeled with the help of the source term, see eq. (\ref{e17c}), which is
determined by the two frequencies, $\Omega_1$ and $\Omega_2$. The effect of $\Omega_2$ is to produce relatively small amplitude short timescale variations,
having the frequency $\approx 2\Omega_{E}^b$ given by eq. (\ref{frecb}). Similar variations are also
caused by the time dependent part of $\Omega_1$ caused by the apsidal precession of the binary orbit.
In order to clarify physical origin of different effects, we consider two
types of numerical runs: 1) 'stationary' runs, where $\Omega_1$ is taken at a particular moment of time (for $\chi=0.5$, its dependency on $r$ is shown in Fig. \ref{Fig1}, and $\Omega_2$ is artificially set to zero. Dynamical variables obtained from
these runs qualitatively characterize those obtained after averaging over times $\gg {\Omega_{E}^{b}}^{-1}$. 2) Fully dynamical
runs, where both $\Omega_2$ and the time evolution of $\Omega_1$ are taken into account. These runs are labeled as $(st)$ and
$(dn)$, respectively. In order to illustrate the difference between the $(st)$ and $(dn)$ variants
of the model, we plot time dependencies of $W_1=\Re[{\bf W}]$ and $W_2=\Im[{\bf W}]$ at $r = 40\,r_g$
as well as their Fourier spectra in Fig. \ref{Fig3}. 
As seen from this Figure, the curves describing the $(st)$ model are
practically periodic, with the period corresponding to the Lense-Thirring precession of the binary.
We check that this is also the case for an extended time span corresponding to the relaxation to the quasi-stationary state.
The curves describing the $(dn)$ model exhibit the same periodicity on the large time intervals, but additionally have superimposed oscillations with a much smaller period.
These oscillations are more pronounced for the curve describing $W_2$. The Fourier analysis shows that this time period coincides
with $\pi/\Omega_{E}^{b}$ \footnote{It is interesting to note that there is another well pronounced peak in the spectrum of the $(dn)$ model corresponding to two times smaller period $\pi/(2\Omega_{E}^{b})$.}.

We use $\tau =\Omega_{E}^b t$ as a dimensionless time variable, starting from the zero initial inclination. We
typically numerically evolve our dynamical system
up to one thousand $\tau$ to ensure its relaxation to the quasi-stationary state.  The details of our fully relativistic disk model and the numerical approach are provided in Section 4.2.1 of \cite{ZI}.
We then plot the inclination angle, $\beta$,
and additionally the wave-like contribution to inclination angle, $\beta_{wave}$, in the case of thin disks, comparing it with the analytic results obtained above.
In all shown runs we use rather small $\beta_b=0.1$, but we have checked numerically
obtained ratios ${\bf W}(t,r)/\beta_b$ remain qualitatively the same when $\beta_b$ is changed.

\subsection{The thin disk models}
\label{thin_models}

\begin{figure}
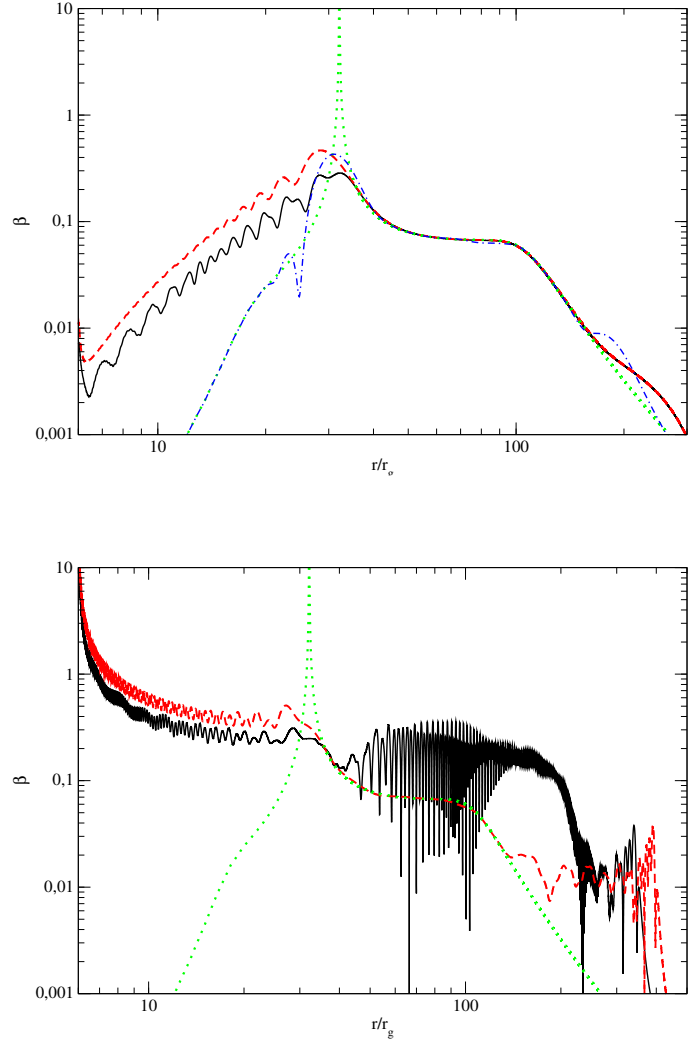

    \includegraphics[width=0.5\linewidth]{beta_half.eps}

    \vspace{1.cm}

    \includegraphics[width=0.5\linewidth]{beta_lalf.eps}
\caption{Top panel.
The dependencies $\beta$ on $r/r_g$ are shown for the time $840 \tau$ in the case of thin disk,
$\delta = 2\cdot 10^{-3}$. Curves on the top panel represent
the case of relatively large viscosity $\alpha = 2\cdot 10^{-2}$, while
curves on the bottom panel represent the case of relatively small viscosity $\alpha = 2\cdot 10^{-4}$. Solid (black) and dashed (red) curves represent
the $(dn)$ and $(st)$ solutions, while the dotted (green) curves show the
'equilibrium' solution (\ref{equ1}). We additionally show the $(st)$ solution obtained for the large viscosity $\alpha = 0.2$ represented by the dot-dashed (blue) curve.}
\label{Fig4}
\end{figure}

\begin{figure}
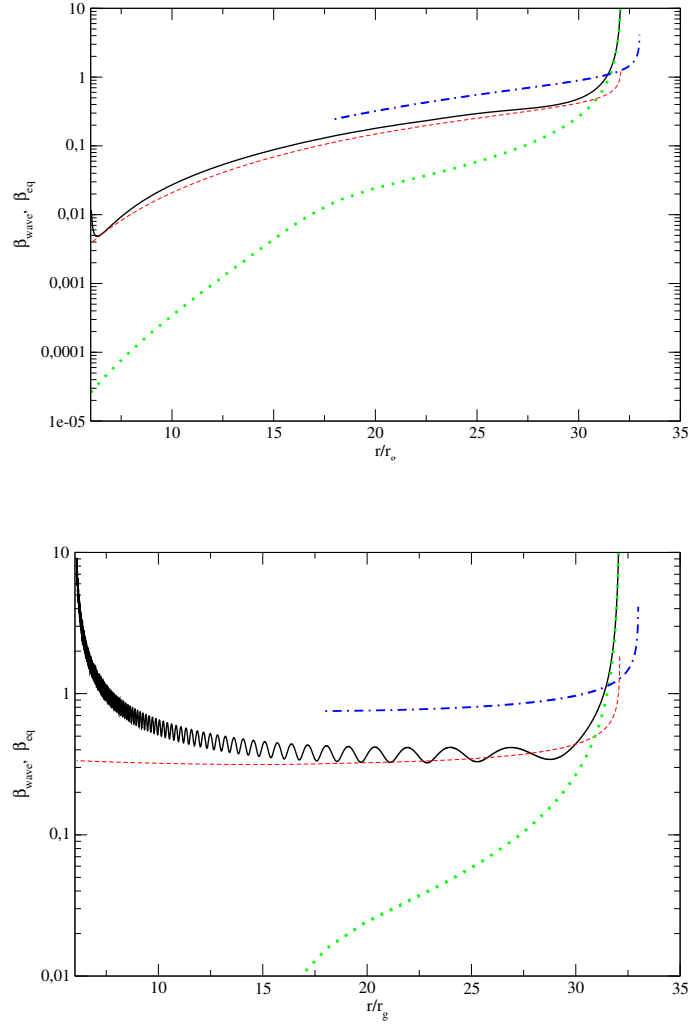

    \includegraphics[width=0.5\linewidth]{betawave_0_02.eps}

    \vspace{1.cm}

\includegraphics[width=0.5\linewidth]{betawave_0_0002.eps}
\caption{The absolute value of the wave-like part of ${\bf W}$, $\beta_{wave}$, is plotted as a function of $r/r_g$ in the case of a thin disk,
$\delta=2\cdot 10^{-3}$. Only the range of radii $r < r_r$ is shown. Top and bottom panels represent the cases
$\alpha =2\cdot 10^{-2}$ and $2\cdot 10^{-4}$, respectively. Solid,
dashed, dot-dashed and dotted (black, red, blue and green) lines show the numerically obtained
$(st)$ curve for the same moment of time as in Fig. \ref{Fig4},
the WKBJ solution (\ref{ne14n}), the analytical expression given
by eqns (\ref{nbeta3}) and (\ref{phi}), and the 'equilibrium' solution
(\ref{equ1}), respectively.
As seen from the bottom panel, in the case of the low viscosity disk, $\beta$ does not tend
to zero when $r$ gets smaller, and the inner part of this disk does not align with the equatorial plane.
Accordingly, the standard Bardeen-Petterson effect is absent, as was shown analytically (see e.g. \cite{II}
and \cite{ZI}) and numerically (see e.g \cite{FB} and \cite{Zhur}) for low viscosity prograde $\chi > 0$
twisted disks.}
\vspace{1cm}
\label{Fig5}
\end{figure}

We show numerically obtained profiles of $\beta(r)$ for the
case of small $\delta=2\cdot 10^{-3}$ in Fig. \ref{Fig4}. The top and
bottom panels correspond to $\alpha=2\cdot 10^{-2}$ and $2\cdot 10^{-4}$, respectively.
As seen in these panels, when $r < r_r \approx 30r_g$, there is an
excess of the amplitude, $\beta$, in comparison with the 'equilibrium'
solution described by eq. (\ref{equ1}) at $r\sim r_r$. This excess is
due to the effect of launching the ingoing wave of twist at the
resonance discussed in Section \ref{wkbj}. For a reference, we additionally plot $\beta(r)$ describing the thin disk with the large viscosity, $\alpha=0.2$,
which was considered in detail by \cite{IZ}. Clearly, in the latter case the resonant production of twist is fully suppressed, and the numerical profile perfectly recovers the 'equilibrium' solution (\ref{equ1}) inward from the resonance.
When $r > r_r$, the 'equilibrium' solution describes the form of $\beta(r)$ quite well for all numerical
cases apart from the $\alpha=2\cdot 10^{-4}$ $(dn)$ case, where strong
radial oscillations of $\beta$ are observed at the scales $\sim 40-200r_g$. We have checked that the radial scale of this oscillations
is much larger than the grid size, and this effect does not seem to be related
to any numerical artifacts. However, these oscillations, as well as oscillations of $\beta$ at $r < r_r$ caused
by relativistic effects (see \cite{II}, \cite{ZI} and references therein)
must be destroyed due to the instabilities present in strongly oscillating low viscosity twisted disks, see e.g. \cite{OL}.
Therefore, small scale oscillations of $\beta$ appear to be physically unreasonable, and we do not discuss them in this work.

In Fig. \ref{Fig5} we compare the wave-like part of the disk inclination obtained numerically with what is given
by eq. (\ref{ne14n}) and eqns (\ref{nbeta3}, \ref{phi}) as well as the equilibrium part of the tilt given by eq. (\ref{equ1}), for
$\delta=2\cdot 10^{-3}$, $\alpha=2\cdot 10^{-2}$ and $\alpha=2\cdot 10^{-4}$. We note that
in order to obtain the wave-like part of the signal, we make a rotation of the numerically obtained profile
${\bf W}(r)$, ${\bf W}(r) \rightarrow  e^{i\gamma_{eq}}{\bf W}(r)$, by a constant angle $\gamma_{eq}$,
and choose this angle by the requirement
that $\Im \left [e^{i\gamma_{eq}}{\bf W}(r)\right ]$ vanishes at a given value of $r > r_r$ sufficiently
far from the resonance.
This procedure approximately corresponds to the change ${\bf W} \rightarrow \hat {\bf W}$. We then calculate $e^{i\gamma_{eq}}{\bf W}(r)-\beta_{eq}$ and use the absolute value of the result,
which is identified with a numerically obtained  $\beta_{wave}$. We see that in the region
 $r < r_p$ our expression (\ref{ne14n}) has an excellent agreement with the data.
When $\alpha=2\cdot 10^{-2}$ the agreement is, in fact, quite good for all radii $r < r_r$.
In the case of $\alpha=2\cdot 10^{-4}$ the numerical curve deviates at small radii. It is clear that at these radii the solution is close to the form of eq. (\ref{th1}) and one can describe the data in a better way by matching
eqns (\ref{th1}) and (\ref{ne14n}) at some
intermediate radius. However, the sharp spatial oscillations of $\beta_{wave}$ at the small radii lead to instabilities
(see, e.g. \cite{OL}, \cite{HO} and references therein) and the solution at the small radii cannot be physical. Our expressions
(\ref{nbeta3}, \ref{phi}) show deviation from the numerical solutions by a factor of two. But they are fully analytical and can be
used for an estimate of the amplitude of the wave-like contribution without any numerical calculations in all appropriate systems, where the value of $\delta$ is sufficiently small.

\subsection{The relatively thick disk models}
\label{thick_models}

Let us now discuss the radial variations of $\beta$ in the relatively thick disk
models with $\delta=0.2$ and $0.5$. Additionally, we consider the case of an
'intermediate' $\delta=0.05$.

\begin{figure}
    \includegraphics[width=0.5\linewidth]{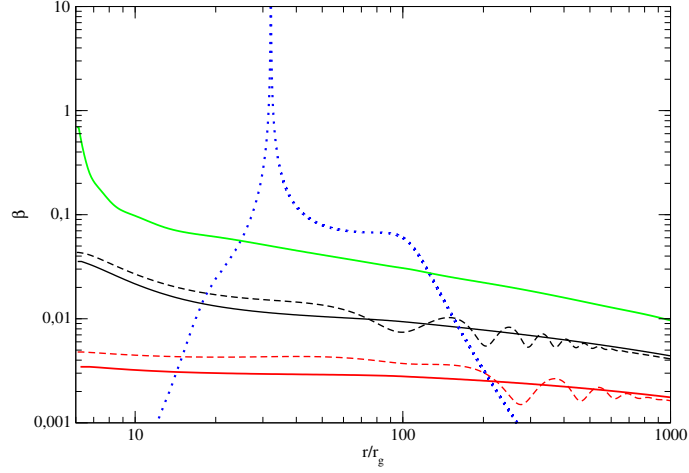}
\caption{
The dependencies $\beta$ on $r/r_g$ for our models of
thick disks. We show the results of calculations based
on the $(st)$ models with $\delta=0.5$, $0.2$ and $0.05$ as solid lines
of different color. Curves with larger values of $\beta$ correspond to
smaller values of $\delta$. Additionally, for $\delta=0.5$ and $0.2$
we show the results of calculations based on the $(dn)$ models as
dashed lines. The dotted line represents
solution (\ref{equ1}).
}
\label{Fig6}
\vspace{0.5cm}
\end{figure}

We show the profiles of $\beta(r)$ in Fig. \ref{Fig6}
for both $(st)$ and $(dn)$ models with
$\delta=0.2$, $0.5$ and $0.05$, $\alpha=2\cdot 10^{-2}$
and the time $\tau=840$ for all cases.
Comparing Figs. \ref{Fig4} and \ref{Fig6} we see that the
profiles corresponding to the thin and thicker disks are quite different. 
Unlike the thin disk case, the curves of $\beta(r)$  corresponding to the thicker disks are much more featureless.

\begin{figure}
    \includegraphics[width=0.5\linewidth]{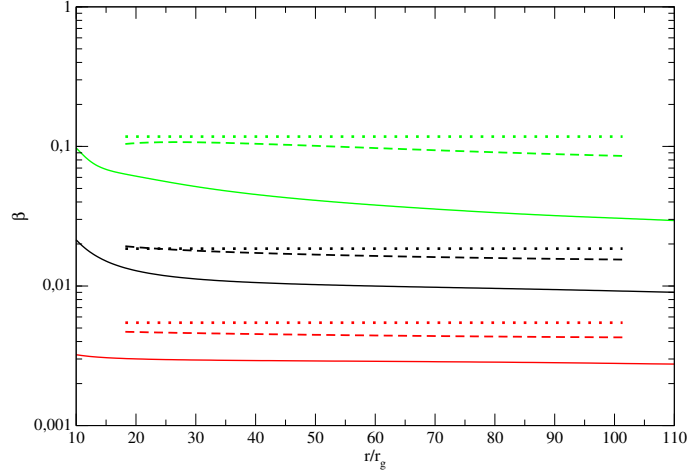}
\caption{
Solid curves represent the same numerical
values of $\beta$ as plotted in Fig. \ref{Fig6} for $(st)$ models with $\delta=0.5$,
$0.2$ and $0.05$ shown in the range of
radii $r_p < r < r_a$. Larger values of $\beta$ correspond
to smaller values of $\delta$. Dashed curves and dotted lines
represent the analytic solution (\ref{fat5})
and the estimate (\ref{fat6}), respectively.
}
\label{Fig7}
\end{figure}

The important features of the thin disk model, such as the presence of an 'equilibrium' solution at $r > r_r$ and the shape of the disk in the vicinity of $r=r_r$ and at $r < r_r$ determined
by the resonant launching of the wave-like contribution are absent in the thicker disk case. It is also important that typical values
of $\beta$ are much smaller than $\beta_b$, with smaller values of $\beta$ corresponding to the larger values of $\delta$.
Fig. \ref{Fig7} shows the profiles of $\beta(r)$ within the orbit together with analytic
curves obtained from eq. (\ref{fat5}). It is seen that the analytic
theory is in a qualitative agreement with the numerical results, as they differ by a factor order of $1.5$
for the cases with $\delta=0.2$ and $0.5$ and by a factor $\sim 3$ for the case $\delta =0.05$. Note that the latter
case lies on the boundary of the applicability of the assumptions adopted in Section \ref{thick},
and, therefore, it's not surprising that the agreement is only qualitative in
this case. Also, we note that the variable $y$ defined in eq. (\ref{th1}) is larger
than unity in this case, and, therefore, the expression (\ref{fat1}), which is valid only
when $y$ is small, is not applicable.

\subsection{A number of intersections between the perturber and the disk per one orbital period in the different
disk models}
\label{intsns}

\begin{figure}
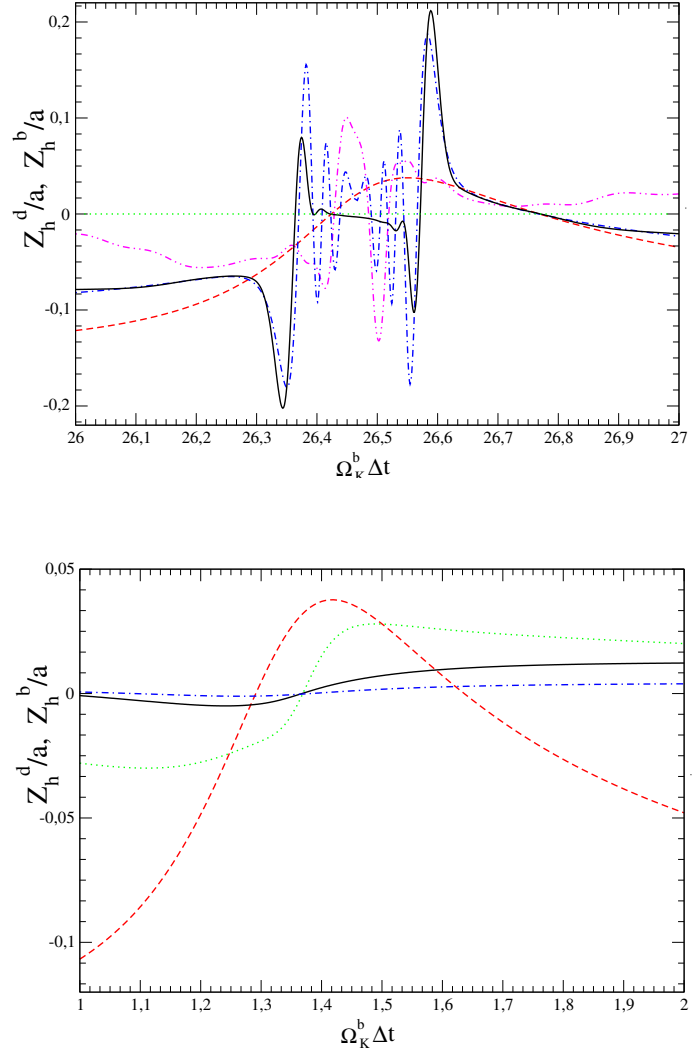

    \includegraphics[width=0.5\linewidth]{intsn_thin_ff.eps}
         \vspace{1.cm}

    \includegraphics[width=0.5\linewidth]{intsn_thick_ff.eps}
\caption{
Top panel. The dashed (red) curve shows the height of the perturber, $Z_h^b$, above the equatorial plane of the primary black hole as a function of time;
solid (black), dot-dashed (blue) and dot-dot-dashed (magenta) curves show the simultaneous height of the disk above the equatorial plane of the primary black hole at the location of the perturber, $Z_h^d$,
for thin disk with $\delta=2\cdot 10^{-3}$ and $\alpha=0.2$, $2\cdot 10^{-2}$ and $2\cdot 10^{-4}$, respectively. Both quantities are expressed in units of $a$. The notations are taken from \cite{IZ}.
Additionally, the dotted line $Z_h=0$ shows the equatorial plane of the primary black hole.
Bottom panel. The same as the top panel, but for a thicker disk.
Solid (black), dot-dashed (blue) and dotted (green) curves represent $\delta=0.2$, $0.5$ and $0.05$, respectively. In all cases,
$\alpha=0.02$. Note that $\Delta t=0$ corresponds to the time $840 \tau$.
}
\label{Fig9}
\end{figure}

It was shown in \cite{IZ} that the presence of the twisted disk leads to multiple crossings of the disk by the secondary black hole per one orbital period.
The spiral structure of the twist found in the low viscosity thin disk solutions, see Figs. \ref{Fig4} and \ref{Fig8}, makes this conclusion even more inevitable.
We confirm it using the technique described in Section 4.3 of \cite{IZ}. For that, we consider the distance
from the secondary and from the disk midplane to the equatorial plane of the primary, denoted as $Z_h^b$ and $Z_h^d$, respectively. Both distances are measured along the same line perpendicular
to the equatorial plane of the primary, as functions of time. Since, by definition, other coordinates of the secondary and the disk midplane coincide with each other, the secondary intersects the disk when $Z_h^b=Z_h^d$.
The functions $Z_h^b(t)$ and $Z_h^d(t)$ for the low viscosity thin disks are shown on the top panel in Fig. \ref{Fig9}.
Only $(st)$ solutions are used. Note that $(dn)$ solutions produce qualitatively the same picture apart from variation of times of intersections from one period to another due to
the non-stationary disk evolution along with the evolution of the binary orbit.
There are formally 6 (the same as in Fig. 9 of \cite{IZ}), 16 and 8 crossings in the case of $\alpha=0.2, 2\cdot 10^{-2}$ and $2\cdot 10^{-4}$, respectively.
These numbers of intersections should be compared with those corresponding to the intersections of the secondary
with the flat disk, $Z^b_h=0$. From the Figure it is clear that there are only two moments of time when  $Z^b_h=0$
per one orbital period. The situation changes dramatically when the intersections with the thick disk are considered.
We see that the ordinary two crossings of the disk by the secondary black hole per one orbital period take place,
see the bottom panel in Fig. \ref{Fig9}.
Obviously, the thick disk with $\delta \gtrsim 0.2$ is too close to the equatorial plane of the primary black hole to produce
more than two crossings by the secondary and, moreover, any significant deviations from the time of intersections corresponding to the flat disk case.

\subsection{Three-dimensional visualization of twisted disks with different parameters}
\label{3D}

\begin{figure}
\includegraphics[width=1.0\linewidth]{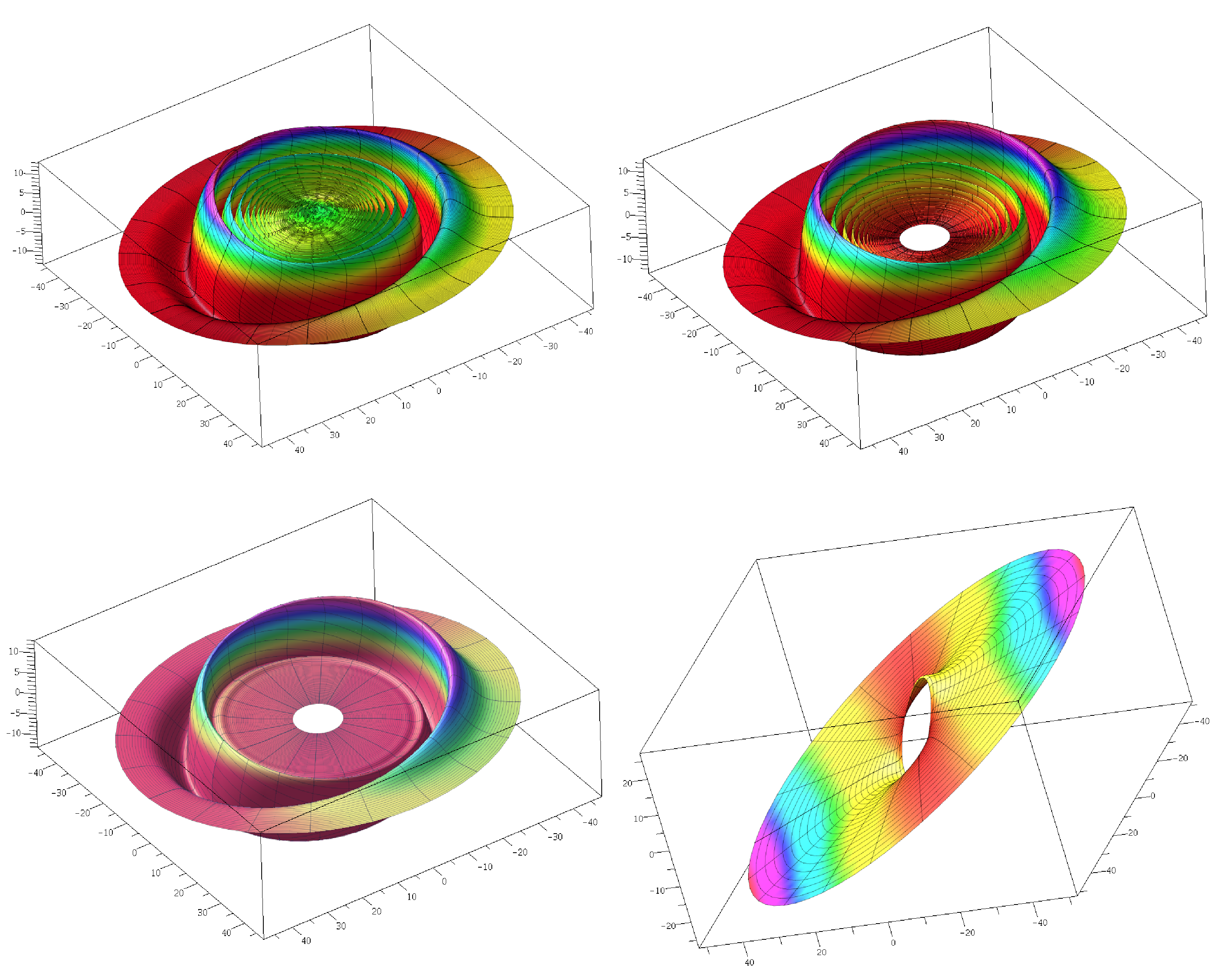}
\caption{Three-dimensional images of the twisted disk represented by its  midsurface (twisted $Z-$coordinate equal to zero, see \cite{IZ} for its precise definition) in the Cartesian coordinates associated with the primary black hole, where the plane $Z=0$ corresponds to its equatorial plane.
The $Z$-coordinate along the surface encodes color in order to render the surface. 
Here are used the $(st)$ solutions, which are also illustrated in Figs. \ref{Fig4} and \ref{Fig6} above. Top-left and top-right panels show the thin disk, $\delta=2\cdot 10^{-3}$, with viscosity $\alpha=2\cdot 10^{-4}$ and $2\cdot 10^{-2}$, respectively.
For a comparison, the bottom-left panel shows a twisted thin disk with the large viscosity $\alpha=0.2$. The bottom-right panel shows the thicker disk with $\delta=0.2$ and $\alpha=0.02$. In the latter case, the value of $Z$-coordinate has been multiplied
by the factor $50$ in order to resolve the disk's tilt and twist.
Note that the observer's line of sight for the thicker disk differs from that of the thin disks for the purpose of
a better illustration of the former case.
All images correspond to $\tau=840$.}
\label{Fig8}
\end{figure}

The three-dimensional images of the inner part of the thin disk are shown in Fig. \ref{Fig8} for different values
of  viscosity. The parts of the disk at radii $r > r_r$ are not shown in order to resolve better the inner
parts, where the twist wave propagates in the case of the disks with $\delta \ll 1$.
For simplicity, we also restrict ourselves by showing only the $(st)$ solutions.

The case of $\alpha=0.2$, where the twist wave is absent (for details, see \cite{IZ}),
should be compared to two other cases of lower viscosity considered in this work. In the absence of the twist wave, one finds that the disk is mainly disturbed close to the resonance. The part of the disk inward to the resonance
is aligned with the equatorial plane of the primary black hole. When considering the smaller  viscosity, $\alpha=2\cdot 10^{-2}$, the inner part of the disk adjacent to the resonance turns into the tight twisting spiral.
Its amplitude vanishes when
the inner edge of the disk is approached.
When the lowest viscosity, $\alpha=2\cdot 10^{-4}$, is considered, the twisting spiral remains undamped at
all radii $r < r_r$ as discussed above.

The three-dimensional image of the disk with $\delta=0.2$ and $\alpha=0.02$ is shown in the bottom right panel of Fig. \ref{Fig8}. Its shape is quite different from all thin disk cases, it is much flatter.
Since the resonance is inefficient in this case, the thick disk is relatively weakly twisted and
tilted with respect to the equatorial plane of the primary black hole. Note that in order to visualize its non-planar shape, we artificially expand the $Z-$coordinate along the rotational axis of the primary  by $50$ times.

\section{Discussion and conclusions}
\label{discussion}
\subsection{The possibility of disk breaking and non-linear behaviour at finite orbital inclinations}
\label{nonlinear}
As we have mentioned above, our equations describing the disk tilt and twist were derived under the assumption that the disk inclination $\beta$ is small, and they may become qualitatively invalid when $\beta_b$ is large enough. In particular, low
viscosity twisted disks may be susceptible to disk breaking. As a criterion of a qualitatively different behavior we
use the condition that
\begin{equation}
\left | r{\partial \over \partial r}{\bf W} \right | > 1,
\label{nl}
\end{equation}
see e.g. \cite{HO} for a recent discussion and appropriate references. We analyze the condition (\ref{nl}) separately for
the thin and relatively thick disks.

For the thin disks, it is seen from, e.g., Fig (\ref{Fig4}) that the main contribution to the radial gradient of ${\bf W}$ 
within the orbit appears
to be determined by the wave-like part $\hat {\bf W}_{wave}$ at the radii smaller than $r_r$ (we neglect the strong oscillations
of $\beta$ at $r > r_r$ in the low viscosity $(dn)$ case, which do not seem to be physical). In order to estimate this gradient, we use (\ref{ne13n}) and substitute the resulting expression in (\ref{nl}), thus obtaining
\begin{equation}
\left | r{\partial \over \partial r}\hat {\bf W}_{wave} \right | \approx \sqrt{\pi}\, { {\Delta \mbox{\boldmath$\omega$}}^{1/4} \Omega^{(r)}_1{\bf \Omega}_c^{1/2}\over \Omega^{(r)}_{s}{\Omega_{res}}^{1/2}},
\label{nl1}
\end{equation}
We then substitute (\ref{q2}) and (\ref{nbeta2}) in (\ref{nl1}), neglecting ${\tilde r}^3$ on the r.h.s. there to have
\begin{equation}
\left | r{\partial \over \partial r}\hat {\bf W}_{wave} \right | \approx {\sqrt{\pi}\,2^{3/2}3^{1/2}\over \delta^{3/2}} 
{\left ({r_g\over r_r} \right )}^{3/4} \,
{\tilde \Omega_{res}}^{1/4}\,q.
\label{nl2}
\end{equation}
We use in (\ref{nl2}) our 'standard' values $r_r/r_g \approx 30$, $\tilde \Omega_{res}\approx 1.6\cdot 10^{-2}$ and $q\sim 
10^{-2}$, and 
substitute it in (\ref{nl2}) to obtain the condition
\begin{equation}
\beta_b \lesssim 4\cdot 10^{-2}\delta^{3/2}. 
\label{nl3}
\end{equation} 
The condition (\ref{nl3}) tells us that the razor-thin disks with $\delta \sim 10^{-3}$ can experience the disk breaking instability.
We note, however, that these disks may have some extended atmosphere, which may contribute to their stability.

\begin{figure}
\includegraphics[width=0.5\linewidth]{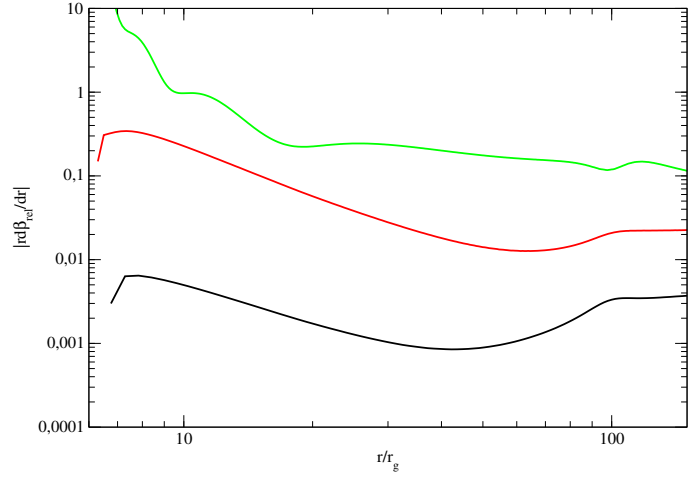}
\caption{The dependencies of $|r{\partial \over \partial r} \beta_{rel} |$ on $r/r_g$ calculated for our thicker disk models with $\delta=0.5$, $0.2$ and $0.05$. Curves with larger values of this quantity correspond to smaller values of $\delta$.}
\label{Fig10}
\end{figure}

As follows from the discussion in Section \ref{thick} and our numerical simulations, see Fig. \ref{Fig6}, the thicker disks have a
rather uniform distribution of $\beta$ within the orbit. Therefore, it is expected that the disk breaking instability is
much less significant in this case. In order to demonstrate it, we introduce the relative disk inclination angle $\beta_{rel}
\equiv \beta/\beta_{b}$ and rewrite the condition (\ref{nl}) as 
\begin{equation}
\left | r{\partial \over \partial r}\beta_{rel} \right | > \beta^{-1}_b,
\label{nl4}
\end{equation} 
where we neglect variations of the angle $\gamma$ assumed to be small for the low viscosity disks considered in this Paper.
The result of the numerical calculation of $|r{\partial \over \partial r}\beta_{rel}|$ is presented in Fig. \ref{Fig10}.    
Note that in order to calculate $\beta_{rel}$, we divide the obtained values of $\beta$ by $\beta_b=0.1$ used in this study,
assuming that $\beta_{rel}$ is qualitatively the same for different values of $\beta_b$.  As seen from this Figure, the gradient
of $\beta_{rel}$ within the orbit is quite small even in the marginal case $\delta=0.05$. Say, in this case  we have
$|r{\partial \over \partial r}\beta_{rel}|(r\approx r_{a})\approx 0.2$. We conclude that for our thicker disk models, the breaking
instability appears to be insignificant.

\subsection{A possible physical state of the accretion disk in the PM model and an estimate of $\delta$}
\label{estimates}
That the secondary collides with the disk a few times per its period could significantly influence its
structure. The collisions result in additional heating of the disk through shocks formed after
each collision event (see e.g. \cite{IIN}). Additionally, some part of hot disk gas pushed
from the disk could fall back and mix with the rest of the disk gas.   
We assume that after each
collision the mass of the hot gas is of the order of $\sim r_{acc}^2\Sigma \sim q^2r^2 \Sigma$
(e.g. \cite{IPP}), where
$\Sigma $ is a typical disk surface density, and $r$ is some characteristic radius, which is fixed
to be equal to $100r_g$ from now on 
\footnote{ We follow \cite{IZ}, who argued that, in the case of OJ 287, 
these collisions do not result in the formation of a gap or some cavity in the disk. This is related to a relatively fast 
orbital evolution of the binary black hole due to the emission of gravitational waves. Thus, $\Sigma$ represents some typical 
disk density at the radii order of the binary semi-major axis.}. This gas has energy per unit of mass of the order of the
orbital energy $\sim {GM\over r} \approx 0.01c^2$,
and it is assumed that a fraction $\xi$
of this energy is  effectively transferred to the disk. This process is important when the corresponding rate of the
mass transfer of the hot gas to the disk $\dot M_{Coll} \sim \xi r_{acc}^2\Sigma \Omega_{orb}$
is larger than the mass flow through the disk, $\dot M_{d}$. From the condition $\dot M_{Coll} > \dot M_{d}$ we obtain
\begin{equation}
\xi \gtrsim \alpha \delta^2 q^{-2}\sim 100\alpha_{*} \delta^{2},
\label{di1}
\end{equation}
where $\alpha_{*}=\alpha/10^{-2}$ and we set $q=10^{-2}$ from now on.
We assume that $\dot M_{d} \sim M_{d}/t_{\nu}$, where $M_{d}$ is a typical
disk mass within the region of interest, $M_{d} \sim \Sigma r^2$, and $t_{\nu}$ is a typical
'viscous' evolution timescale, $t_{\nu} \sim \alpha^{-1}\delta^{-2}\Omega_{orb}^{-1}$. Equation (\ref{di1})
tells us
that, for a standard razor-thin disk with $\delta \sim 10^{-3}$, the fraction $\xi $ can be as small
as $10^{-4}$ for the effect of the additional disk heating to be significant.

We also note that the heating rate due to black hole-accretion disk collision
was estimated  in the recent
paper \cite{Num}. This results in $\dot M_{Coll}$ being $\delta^{-1}$ times larger than our estimate.
The discrepancy is determined by their neglect of the fact that the secondary may be considered as
efficiently interacting with the disk only when it crosses it, which takes a time order of $\delta P_{orb}$.
Taking into account this factor and averaging over several periods, one can obtain our estimate. Additionally, it is important to point
out that it is wrong to use the viscous timescale (defined as infall time in \cite{Num})
as a characteristic timescale of a change of angular momentum
in a twisted disk as was done in \cite{Num}. It is well known that when $\alpha > \delta $
the characteristic timescale is $\alpha^2$ times smaller than the viscous timescale, see \cite{PP}. In the opposite
case of $\delta > \alpha $, the characteristic timescale can be obtained from the viscous timescale by a formal substitution
$\alpha=\delta^{-1}$, it is of the order of the sound crossing time, see \cite{LP}.

To proceed any further, we should be able to estimate the mass flow $\dot M_{d}$. In the framework
of the PM model, it is natural to assume that the disk contributes a significant fraction of the total
radiation flux of OJ 287. In this case, its luminosity is expected to be the order of $10^{48}{\rm erg/s}$,
which corresponds to $\sim 5\cdot 10^{-3}L_{Edd}$ (see e.g., \cite{ufn} and \cite{VG}),
where $L_{Edd}={4\pi c\over \kappa_{T}}GM$ is the Eddington luminosity, and $\kappa_{T}$ is the Thomson opacity
We also assume that $\sim $ 6 percent of the rest mass is converted to radiation (e.g., \cite{SS}), thus obtaining $\dot M_d \sim {GM\over c\kappa_{T}}$.
For the accretion rate,
it immediately follows from eq. (2.17) of \cite{SS} that the disk is expected to be
radiation dominated at radii $\lesssim 10^2\,r_g$ even when the external heating is neglected. When this
heating is taken into account, the ratio of radiation pressure to the gas pressure should be even
larger. In what follows, only the radiation pressure is taken into account.

When an $\alpha-$disk is radiation pressure dominated, it may be susceptible to thermal instability
\cite{SS1}. It has been known for a long time that in the case when this instability operates in the disk, it should
be slim with $\delta \sim 1$ \cite{Marek}. However, recent numerical results do not provide any convincing evidence for the existence of a vigorous thermal instability
(see e.g., \cite{Ross}, \cite{Blaes} and references therein) and we assume below that it is inefficient.

Assuming that $\dot M_d \sim {GM\over c\kappa_{T}} \sim  M_{d}/t_{\nu}$
\begin{equation}
\tau_{T}\equiv \kappa_{T}\Sigma \sim \alpha^{-1}\delta^{-2}\left ( {v_{orb}\over c} \right ),
\label{di2}
\end{equation}
where $\tau_{T}$ is the optical thickness with respect to Thomson scattering,
$v_{orb}=r\Omega_{orb}$, and from our assumptions it follows that $v_{orb}/c \sim 0.1$.

On the other hand, assuming that equation (\ref{di1}) is satisfied by a wide margin and that
the heat advection does not dominate the energy balance
in the disk, the radiation flux from the disk surface, $F_r$,
should be of the order of the energy flow of hot gas to the disk per unit of its surface,
$\xi \, ({GM\over r})\, q^2\Sigma\, \Omega_{orb}$. In its turn, from the equation of radiation
transfer in the vertical direction, it follows that the radiation pressure $P_{r} \sim F_r\, \tau_{T}/c$.
We also use the equation of hydrostatic balance in the vertical direction in the form
\begin{equation}
\delta \sim {r^2P_r\over GM\Sigma},
\label{di3}
\end{equation}
and substitute $P_{r}$ expressed in terms of the energy flow of hot gas to this equation.
In this way we obtain
\begin{equation}
\delta \sim \xi q^2 \left ( {v_{orb}\over c} \right )\, \tau_{T}.
\label{di4}
\end{equation}
Now we use eq. (\ref{di4}) to express $\tau_{T}$ in terms of $\delta$ and substitute the result
in eq. (\ref{di2}) to obtain
\begin{equation}
\delta \sim \xi^{1/3}\alpha^{-1/3}q^{2/3}\left ( {v_{orb}\over c} \right )^{2/3}\approx 5\cdot 10^{-2}\xi^{1/3}
\alpha_{*}^{2/3},
\label{di5}
\end{equation}
where we remind that we use $q=10^{-2}$ and $\alpha_{*}=\alpha/10^{-2}$. Equation (\ref{di5})
tells us that unless $\xi$ isn't very small, we expect $\delta \sim 0.1$ in the framework
of the simple model considered above. Note that although the disks with such typical values
of $\delta$ are expected to have inclinations order of $\beta_b$, from (\ref{ne17n}) it follows
that the resonance considered in this Paper shouldn't play a significant role, and the strong
response of the disk's inclination to the presence of the secondary is not expected.
In this situation, only two crossings of the disk per one orbital period are likely to take place, and
we have checked numerically that it is indeed the case. It is also interesting to note that
the viscous time, $t_{\nu}$, is of the order of a few tens of thousands of years, which is in its
turn of the order of the evolutionary time of the orbit due to emission of gravitational waves,
see \cite{IZ}. This makes our model self-consistent, since the disk surface density
has enough time to evolve from larger values corresponding to a disk unperturbed by
the presence of the secondary to the values following from eq. (\ref{di2}).

Substituting eq. (\ref{di5}) into eq. (\ref{di2}) and using the nominal value of the parameters,
we see that in our model the Thomson optical thickness $\sim 10^3$ at $r \sim 10^2 r_g$. Also, it follows
from, say, eq. (\ref{di3}) that in our model the disk temperature should be $\sim 10^{4}K$ at the same radii.

\subsection{Possible extensions of our work}

It would be of interest to compare the spectrum of OJ 287 to the radiation spectrum of the disk model considered above
with the temperature and optical thickness extrapolated from our estimates corresponding
to the radii $r \sim 10^2r_g$  to smaller radii.

Apart from the spectral modelling, it is of interest to develop a more rigorous
variant of the flat disk model along the lines outlined in the previous Section
and use it as a background state for our models of the dynamics of the disk tilt and twist.
Another straightforward
development would be to take into account that the binary orbital parameters evolve at the timescale $\sim 10^4yr$
due to emission of gravitational waves \footnote{Note that such a short timescale of orbital evolution
presents a difficulty for a theory of formation of systems with the parameters appropriate
for the PM model of OJ 287. Indeed, since the evolution timescale is so short, a number of such systems observed in the
present time is expected to be quite small.}.

The important way of extending our results is to try to reproduce them
in a fully numerical approach. It appears
to be a much simpler task for the relatively thick disks with $\delta \sim 0.1$ than for the razor-thin ``standard'
disks. In particular, the double averaging procedure used in \cite{IZ} to determine the gravitational influence of
the secondary on the disk in terms of the frequencies $\Omega_1$ and $\Omega_2$ as described by eq. (\ref{e17c})
must be thoroughly checked.

A preliminary numerical simulation  was made in \cite{Num}, where the authors considered GRMHD simulations
of a binary black hole with the orbital parameters similar to those assumed in the PM model, $\delta= 0.1$ and
a range of mass ratio $q \le 0.1$. Unfortunately, these simulations cannot be directly compared with our
results due to the following reasons. 1) The authors assumed that both black holes are non-rotating. 2)
They computed the evolution of the disk till only $\sim 40\Omega_{b}^{-1}$, which is much smaller
than the expected timescale of relaxation to the quasi-stationary state. 3) They assumed that the orbit
is initially perpendicular to the disk plane. Thus, it is important to extend the results of \cite{Num}
to the case of a binary with the rotating primary with $\chi\sim 0.5$ and somewhat smaller inclination, $\beta_b$.
Additionally, the computational time should be at least one order of magnitude larger.

\subsection{Conclusions}

In this Paper we considered the extension of the model of the twisted disk in systems with the parameters appropriate
for the PM model of OJ 287 proposed recently by us in \cite{IZ}. The important property of the model is the presence
of resonance in the set of equations describing the disk tilt and twist, which leads to a strong deviation of
the disk shape from the flat one. In particular, \cite{IZ} showed that a disk with the relative thickness $\delta=10^{-3}$
and large viscosity parameter $\alpha=0.1$ is strongly twisted within the orbit, with typical inclinations order of
the binary inclination with respect to the equatorial plane, $\beta_b$. In this situation, the secondary
experiences typically $\sim 5-6$ collisions with the disk per one orbital period, which is in tension with the PM model
requiring only two collisions per orbital period.

Here we considered disks with smaller $\alpha=2\cdot 10^{-4}-0.01$
and a range of $\delta$, such as $2\cdot 10^{-3} \le \delta \le 0.5$.
Similar to \cite{IZ} we found that the twisted disk relaxes to a quasi-stationary form in the reference frame precessing
with the Lense-Thirring frequency of the binary orbit, $\Omega_{LT}^{b}$. However, this form is dramatically different
in the cases of thin disks with $\delta \sim 10^{-3}$ and thick disks with $\delta \sim 0.1$.

In the case of the thin disks, the resonance plays a very important role for the whole range of considered $\alpha$.
The disk shape is strongly twisted, while its typical tilt $\sim \beta_b$. Such disks cannot be used in the models
of OJ 287. We found that when viscosity is small enough, a range of radii close to the resonance produces
twisting spiral wave, which is stationary in the precessing frame 
\footnote{It is interesting to note that such a spiral wave
survives until the rather high viscosity $\alpha \gg \delta$. This makes it different from the known radial oscillations of the disk tilt
discovered by \cite{II}, which are suppressed already at $\alpha \sim \delta$.}.
From mathematical point of view, the mechanism of generation
of such waves
is analogous to the standard mechanism of generation of density waves close to Lindblad resonances by satellites
in flat disks, see \cite{GT}, and, e.g., \cite{Kley}, \cite{Paardekooper} for comprehensive reviews.
We developed a WKBJ theory of such waves and showed that it is in a good agreement with our
numerical results. The effect of the generation of waves may find its application in binary black holes containing
thin disks with small viscosity.

When a disk is thick, the resonance is unimportant. In this case the disk inclination doesn't vary much within the binary orbit
and is considerably smaller than $\beta_b$. The systems with such disks have only two collisions of the secondary with the disk per one orbital
period, and, therefore, can be used as the models of OJ 287. We developed a qualitative analytic theory of such twisted disk
and found that it leads to the values of the disk inclination, which are in agreement with the numerical results within
the factor $\sim 1.5-2$. Additionally, we proposed a simple order-of-the-magnitude scenario of the formation of the disks
with $\delta \sim 0.1$ in OJ 287 due to heating of the disk gas by collisions with the secondary.

Since our equations describing the evolution and quasi-stationary configurations of the disk tilt and twist are linear,
it is important to estimate a possible role of non-linear effects. We made such estimates in Section \ref{nonlinear}. From
the results of this Section it follows that while the role of non-linear effects in the razor-thin disk models may be significant
even when $\beta_b$ is small, our thicker disk models are not expected to be significantly influenced by these effects.

We also note that the presence of some extended atmosphere may alleviate the problems experienced by the razor-thin models. Also note that, to the best of our knowledge, all numerical models of the most prominent non-linear effect - the disk breaking—have been done assuming that the disks are locally isothermal; see e.g. \cite{HO}   and \cite{Pepeka}. In our opinion, it is important to consider models that can take into account the expected backreaction of the disk-breaking process on the disk thickness, since it may change the criteria of disk breaking and give some information about a possible final state of a disk undergoing such a process.

The theory discussed above may have some direct observational consequences. At first, it leads to a definite modification
of the spectrum of OJ 287, which can be modeled and compared with the observations. Secondly, it predicts variability
at the large timescale scale $2\pi {\Omega^b_{LT}}^{-1} \sim 10^{3}yr$ and, also, some modulation at somewhat smaller timescale $\pi {\Omega^{b}_{E}}^{-1}\sim
10^2 yr$ (see Fig. \ref{Fig3}), which can, in principle, be observed, for example, through the analysis of the geometrical form
of the jet in OJ 287 at small spatial scales, see e.g. \cite{ufn}.

\bigskip
\section{ }
\label{aps}

\bigskip

\acknowledgements

We are grateful to M. V. Barkov and P. C. Fragile for useful comments. It is our pleasure to thank the referee
for a thorough reading of the manuscript and making a significant number of important comments and suggestions. 

The study was conducted under the state assignment of Lomonosov Moscow State University.

\section{DATA AVAILABILITY}
No data were created or analyzed in this study.




\bibliographystyle{apsrev4-2}
\bibliography{references}

\end{document}